%% file: galois.tex
\documentclass[12pt,a4paper]{article}

\usepackage{amssymb}\usepackage{feynmf}

\begin{document}

\include{qcd}

\def\N{\mathbb{N}}

\newcommand{\be}{\begin{equation}}
\newcommand{\ee}{\end{equation}}
\newcommand{\bea}{\begin{eqnarray}}
\newcommand{\eea}{\end{eqnarray}}
\newcommand{\beas}{\begin{eqnarray*}}
\newcommand{\eeas}{\end{eqnarray*}}

\newtheorem{theorem}{Theorem}
\newtheorem{lemma}[theorem]{Lemma}
\newtheorem{prop}[theorem]{Proposition}
\newtheorem{cor}[theorem]{Corollary}
\newtheorem{defn}[theorem]{Definition}
\newtheorem{rem}[theorem]{Remark}
\newtheorem{conj}[theorem]{Conjecture}
\newtheorem{exam}[theorem]{Example}

\def\build#1_#2^#3{\mathrel{\mathop{\kern 0pt#1}\limits_{#2}^{#3}}}

\def\One{\mathbb{I}}

\title{\bf Anatomy of a gauge theory}
\author{Dirk Kreimer\footnote{\noindent CNRS, also supported in parts by NSF grant DMS-0401262 at
Center for Mathematical Physics, Boston University. Email: kreimer@ihes.fr}\\
{\footnotesize IH\'ES, 35 rte.~de Chartres, 91440 Bures-sur-Yvette,
France}\\}

%
%
\maketitle

\abstract{We exhibit the role of Hochschild cohomology in quantum
field theory with particular emphasis on gauge theory and
Dyson--Schwinger equations, the quantum equations of motion. These
equations emerge from Hopf- and Lie algebra theory and free quantum
field theory only. In the course of our analysis we exhibit an
intimate relation between the Slavnov-Taylor identities for the
couplings and the existence of Hopf sub-algebras defined on the sum
of all graphs at a given loop order, surpassing the need to work on
single diagrams.}

\section{Introduction}
\begin{fmffile}{fmfG-Intro}
Over the last seven years many properties of the Hopf algebra
structure of renormalization have been investigated, mostly on the
mathematical side, with at least one notable exception
\cite{Broadhurst:2000dq} which showed how to solve a non-linear
Dyson--Schwinger equation exactly, as opposed to a mere perturbation
expansion.

In this paper, we explore the elementary relations between a
perturbative expansion in quantum field theory, the corresponding
Hochschild cohomology and the equations of motion in the context of
a generic gauge theory. A major feature underlying our analysis is
the emphasis on free quantum field theory and locality  expressed
through Hochschild cohomology. Together they specify the interacting
theory. The novel phenomenon we report here is the interplay between
the existence of a suitable sub Hopf algebra of perturbation theory
and the existence of internal symmetries: the Slavnov--Taylor
identities which ensure equality of renormalized couplings are
equivalent to the existence of forest formulas indexed just by the
loop number, the grading of the usual Hopf algebra of Feynman
graphs.
\subsection{The structure of Dyson Schwinger equations}
If one ultimately wants to address non-perturbative quantum field
theory one has to solve the corresponding Dyson--Schwinger equations
exactly. This looks prohibitively difficult. But the recent progress
with perturbative quantum field theory also points towards methods
of solutions for Dyson--Schwinger equations
\cite{Bonn,Broadhurst:2000dq}. Let us review the current situation.
Detailed references  can be found in \cite{review}.

Feynman graphs possess a Lie algebra structure which dually governs
the structure of renormalization, via the forest formula. The
corresponding Hopf algebra furnishes one-cocycles which ensure
locality of counterterms. These one-cocycles allow to generate the
one-particle irreducible Green functions and provide the skeletons
underlying the perturbative expansion. The one-cocycles correspond
to primitive elements in the linear span of generators of the Hopf
algebra and are graded by the loop number. One gets a set of
embedded Hopf algebras from a hierarchy of Dyson Schwinger equations
based on primitive graphs up to $n$ loops. Underlying this structure
is a universal Hopf algebra structure whose generators are given by
the sum of all graphs with a given loop number. Let us call this sub
Hopf algebra the grading algebra \cite{Bonn,Bergbauer:2005fb}. It is
the existence of this grading algebra which allows for the
recursions which made the non-perturbative methods of
\cite{Broadhurst:2000dq} feasible.

The purpose of the present paper is to exhibit this structure for
the example of a non-abelian gauge theory. In particular, we exhibit
the grading algebra of this theory and show that its existence is
equivalent to the existence of the Slavnov--Taylor identities for
the couplings.

Dyson--Schwinger equations,  illuminating recursive
structure in Green functions, are recently on the forefront again in
different areas of field theory \cite{Bern,Korchemsky,Serreau}. We
expect that our results are a starting point to understand these
phenomena more systematically. In particular, the representations of
the rescaling group and hence the role of dilatations in quantum
field theory is intimately connected to the role of the Hochschild
one-cocycles above, and motivates our quest for understanding their
role in non-perturbative quantum field theory.
\subsection{Interaction vertices from free quantum fields and locality}
We start with a free quantum field theory with free propagators
given by the usual requirements of free field theory for fermion and
boson fields.

Before we discuss how locality determines the structure of the full
Green functions we look at the tree level first, for motivation. We
want to emphasize in our approach the known fact that the Feynman
rules for tree-level graphs are determined by free quantum field
theory (which gives Feynman rules for the edges) and locality (which
implies Feynman rules for the vertices).

So let us first remind ourselves how the quest for locality
determines the interaction vertices in the context of renormalizable
theories, without reference to a classical Lagrangian at this stage.

We exhibit the argument for the example of the QED vertex. Thus we
aim to infer the local interaction vertex $v_\mu(q,p)$ for the
interaction of a photon, with four-momentum $p-q$, with an $e^+ e^-$
pair, with four-momenta $p$ and $q$.

We will infer the Feynman rule
for the vertex from the knowledge of the free photon, electron and
positron propagator, and from the requirement that we should be able
to renormalize by local counterterms.

We start with an Ansatz that at tree-level the sought-after vertex
must be of the form \be v_\mu(q,p)=\sum_{i=1}^{12} a_i
f^{[i]}_\mu=a_1\gamma_\mu+a_2 \frac{q\!\!\!/
q_\mu}{q^2}+\cdots,\label{treevertex}\ee in accordance with Lorentz
covariance, spin representations  and dimensions of the free
propagators involved. Note that we can assume the coefficients $a_i$
above to be constants: below we are only interested in the behaviour at
large momentum transfer at each internal vertex. In that limit, each ratio in \be
\left\{u=\frac{q^2}{q^2+p^2+(p+q)^2},v=\frac{p^2}{q^2+p^2+(p+q)^2},w=\frac{(p+q)^2}{q^2+p^2+(p+q)^2}\right\}\ee
turns to zero or one, so that $a_i(u,v,w)$ turns to a constant indeed.

We can now determine $v_\mu(q,p)$ as follows.

First, we note that free quantum field theory and the requirement
for renormalizability provide us with powercounting. We hence
construct the 1PI graph with lowest non-vanishing loop order
corresponding to our interaction: $$\qedagffa.$$
This graph is log-divergent. If we demand to have a renormalizable
theory, its coefficient of log-divergence must be proportional to
the sought after tree-level vertex. This is indeed the well-known
clue. By construction, the graph has no subdivergences and hence
will evaluate to an expression of the form \be
\phi\left(\qedagffa\!\right)\sim v_\mu(q,p)\ln \lambda+{\rm finite\,
terms},\ee where we cut-off momentum integrals by $\lambda$. The
important point is that the only part of this vertex functions which
is allowed to diverge with $\lambda$ is the part proportional to the
sought-after tree-level vertex, if we are in a renormalizable
theory.

  Hence, we can write \be v_\mu(q,p) \sim
\lim_{\lambda\to+\infty}\frac{1}{\ln\lambda}\int_{-\lambda}^{+\lambda}d^4k
\left[v_\sigma(q,k)
\frac{1}{k\!\!/+q\!\!/}v_\mu(k+q,k+p)\frac{1}{k\!\!/+p\!\!/}v_\rho(k,p)
D_{\sigma\rho}(k)\right].\ee Inserting (\ref{treevertex}) we
indeed find that \be v_\mu(q,p) \sim \gamma_\mu,\ee as was to be
expected.

 Note that this states the obvious: in a local
renormalizable quantum field theory the short-distance singularities
can be absorbed by suitable modifications of parameters in the given
Lagrangian. Vice-versa, if we make locality the starting point and
hence require that short-distance singularities are self-similar to
the tree-level terms, we obtain the interaction part of the
Lagrangian from this requirement.

There is a cute albeit obvious message hidden here: if we settle on
a given set of free fields and demand that they interact in a
renormalizable manner, already the lowest loop order scattering
processes fix the form of the interaction vertices from the
knowledge of free quantum field theory and the requirement of
locality, whilst the interaction part of the Lagrangian appears as a
derived quantity. In this spirit we will continue to explore what we
can learn about QFT, in particular about a gauge theory, starting
from free covariances and so-determined interaction terms, and the
accompanying one-particle irreducible graphs which go with them. Our
guiding principle will still be locality, in its mathematical
disguise as Hochschild cohomology.

In particular, we are now interested in formal sums of graphs
corresponding to a specific instance of propagation or interaction:
the sum of all 1PI graphs which constitute in perturbative QFT the
1PI Green function for that amplitude. The sum over all such 1PI
Green functions furnishes then the effective action, by definition.
We will study these 1PI Green functions for a non-abelian gauge
theory.

 We hence wish to discuss the formal sums \be
\Gamma^{r}=\sum g^{2|\Gamma|}\frac{\Gamma}{{\rm sym}(\Gamma)}.\ee
Here and in the following the superscript $r$ specifies the Green
function under consideration, it can be regarded as a collective
label for the quantum numbers at the external legs of that function.

Already at this level interesting structure emerges.  Our main tool
will be the exploration of the Hochschild cohomology of the Hopf
algebra structure which comes with such graphs. This Hochschild
cohomology provides a mathematical precise formulation of locality \cite{Kreimer:2002rf,Bergbauer:2004cx,Bergbauer:2005fb},
and will carry us far in the understanding of the structure of the
theory.

\subsection{Quantum equations of motion from Hochschild Cohomology}
Such Hopf algebras have marvelous properties which they inherit from
the universal object ${\cal H}_{\rm rt}$ for such algebras: the
commutative Hopf algebra of non-planar rooted trees
\cite{Connes:1998qv}.

In that context, it is beneficial to study Dyson--Schwinger
equations as formal constructions based on the Hochschild
cohomology of such Hopf algebras. Before we justify the connection
to Dyson--Schwinger equation through the study of a generic gauge
theory below, let us first describe the universal set-up on rooted
trees \cite{Bergbauer:2005fb}.

First, we settle on say a suitable Hopf algebra $A$ which can be
${\cal H}_{\rm rt}$ or any suitable sub-Hopf algebra $A$.

Let $A$ then be any such connected graded Hopf algebra which is free or
free commutative as an algebra, and $(B_+^{d_n})_{n\in\N}$ a
collection of Hochschild 1-cocycles on it.  The Dyson-Schwinger
equation is
\begin{equation}\label{eq:genDSE}
X=\One+\sum_{n=1}^\infty \alpha^n w_n  B_+^{d_n}(X^{n+1})
\end{equation}
in $A[[\alpha]].$ The parameter $\alpha$ plays the role of a
coupling constant. The $w_n$ are scalars in $k.$ We decompose the
solution
\begin{equation}
X=\sum_{n=0}^\infty \alpha^n c_n
\end{equation}
with $c_n\in A.$
\begin{lemma}\label{lem:solgenDSE}
The Dyson-Schwinger equation (\ref{eq:genDSE}) has a unique solution
described by $c_0=\One$ and
\begin{equation}\label{eq:genreccn}
c_{n} = \sum_{m=1}^{n} w_{m}B_+^{d_{m}}\left(\sum
_{k_1+\ldots+k_{m+1}=n-m,\, k_i\ge 0} c_{k_1}\ldots
c_{k_{m+1}}\right).
\end{equation}
\end{lemma}
The $c_n$, coefficients in the $n$-th term of the perturbative
expansion have a very nice property which we will rediscover in
quantum field theory:
\begin{theorem}\label{thm:dsesub}
 The $c_n$ generate a Hopf subalgebra (henceforth called the grading algebra) of $A:$
\begin{equation}
\Delta(c_n)= \sum_{k=0}^n P_{n,k}\otimes c_k \label{eq:subcoeff}
\end{equation}
where the $P_{n,k}$ are homogeneous polynomials of degree $n-k$ in the
$c_l,$ $l\le n:$
\begin{equation}\label{eq:pnk}
P_{n,k} = \sum_{l_1+\ldots+l_{k+1}=n-k}c_{l_1}\ldots c_{l_{k+1}}.
\end{equation}
In particular, the $P_{n,k}$ are independent of the $w_n$ and
$B_+^{d_n}.$
\end{theorem}

In this paper we want to discuss this structure when we pass from
the universal object ${\cal H}_{\rm rt}$ to the concrete Hopf
algebra of say a generic gauge theory. We aim at insights into the
non-perturbative structure of QFT and also prepare for new
methods of computation in subsequent work. In particular we use the fact that the
operadic proof of the above theorem given in \cite{Bergbauer:2005fb}
extends to our case once the proper insertion maps for graph
insertions have been defined. To see the main point, we study first
an elementary example.
\subsection{A toy Ward identity}
Consider the following system of DSEs based on say four Hochschild
one-cocycles. \bea X_1 & = & 1+\alpha
B_+^a(X_3X_1)+B_+^b(X_2^2)\\
X_2 & = & 1+\alpha B_+^c(X_2^2)\\
X_3 & = & 1+\alpha B_+^d(X_3^2). \eea One immediately confirms that
imposing the symmetry \be X_1 X_3=X_2^2 \label{symmtoy} \ee in the
Hopf algebra is equivalent to giving the sub Hopf algebra, $i\in
\{1,2,3\}$, \bea X_i
& = & 1+\sum_n \alpha^n c_n^{(i)}\\
\Delta(c_n^{(i)}) & = & \sum_{k=0}^n P_{n,k}^{(i)}\otimes
c_{n-k}^{(i)}, \eea where the polynomials $P_{n,k}^{(2)}$ and
$P_{n,k}^{(3)}$ are easily determined as in (\ref{eq:subcoeff}).
Similarly, upon using the symmetry (\ref{symmtoy}) we find a new
equation for $X_1$ \be X_1=1+\alpha \left[ B_+^a+B_+^b\right](X_1
X_c),\ee where $X_c:=X_3=X_2^2/X_1$, and all elements $c_n^{(1)}$
are symmetric in exchange of labels $a$ and $b$. The existence of a
sub Hopf algebra on the generators $c_n^{(i)}$ is now
straightforwatd to establish as in \cite{Bergbauer:2005fb}.

We will now focus on the case of a generic non-abelian gauge theory,
and exhibit how the Hochschild cohomology of the Hopf algebra of its
perturbative expansion, the equations of motion and local gauge
symmetry interfere. In particular, we will find a similar situation:
the existence of a sub Hopf algebra is equivalent to the existence
of relations exhibiting symmetries between Green functions.

 To formulate our results we first introduce the pre-Lie algebra of
Feynman graphs in this context in the next section.

We then introduce our result and discuss it with the help of a
completely worked out two-loop example.

\end{fmffile}
\section{Graphs}
In this section we first define graphs and the accompanying pre-Lie
and Hopf algebras. The material is a straightforward application of
previous results to a generic gauge theory.
\begin{fmffile}{fmfG-Exa0}
\subsection{The set of graphs}
All graphs we consider are built from the following set $R$ of
edges and vertices \be R=\left\{
\qcdzff,\qcdzuu,\qcdzggb,\qcdzgff,\qcdzguu,\qcdzggc,\qcdzggd\right\}.\ee
We subdivide into edges and vertices, \be R_{\rm V}=\left\{
\qcdzgff,\qcdzguu,\qcdzggc,\qcdzggd\right\},\ee and \be R_{\rm
E}=\left\{ \qcdzff,\qcdzuu,\qcdzggb\right\}.\ee Obviously,
$R=R_{\rm E}\cup R_{\rm V} $. We have included edges for the free
propagation of the local gauge field, corresponding ghost fields,
and fermion fields as the only matter fields. We exclude the
discussion of scalar matter fields coupled to gauge fields which
deserve a separate discussion in future work. We thus include only
the expected vertices in a generic local gauge theory: triple and
quartic self-interactions of the gauge field, an interaction of
the gauge field with its ghost field and minimal interaction
between the gauge and matter fields - $R_{\rm V}$ is determined by
$R_{\rm E}$ and locality  in the spirit of the argument in the
previous section.

We then define one-particle irreducible (1PI) graphs as usual: they
remain connected after removal of any one of the internal edges. For
such a 1PI graph $\Gamma$ we have external legs $\Gamma^{[1]}_{\rm
ext}$, internal edges $\Gamma^{[1]}_{\rm int}$ and vertices
$\Gamma^{[0]}$.

 For any 1PI graph
$\Gamma$ we let ${\bf res}(\Gamma)$ be the graph when all its
internal edges shrink to a point \be {\bf
res}(\qcdaggb\!)=\qcdzggb\!.\ee We call ${\bf res}(\Gamma)$ the
residue of $\Gamma$ to emphasize that a graph $\Gamma$ provides a
counterterm $S_R^\phi(\Gamma)$ which contributes $S_R^\phi(\Gamma)
\phi({\bf res}(\Gamma))$ to the Lagrangian \be {\cal L}=\sum_{r\in
R_{\rm E}}1/\phi(r)+\sum_{r\in R_{\rm V}}\phi(r),\ee where we note
that for $r\in R_{\rm E}$ we have $1/\phi(r)$ as the inverse free
propagator, as it should be. We extend the notion of a residue of a
graph to a product $\Pi_i(\Gamma_i)$ of graphs: \be {\bf
res}\left(\Pi_i \Gamma_i\right)=\Pi_i{\bf res}(\Gamma_i).\ee

Any element $r$ in $R$ has a superficial degree of divergence (sdd),
$w(r)$, given as follows \bea
w(\qcdzff)=1,w(\qcdzuu)=1,w(\qcdzggb)=2,w(\qcdzgff)=0,\nonumber\\w(\qcdzguu)=0,w(\qcdzggc)=-1,w(\qcdzggd)=0.
\eea  We introduce the loop number $|\Gamma|$, \be |\Gamma|={\rm
rank}(H_1(\Gamma)),\ee where $H_1(\Gamma)$ is the first homology
group of $\Gamma$, and \be
\left|\prod_i\Gamma_i\right|=\sum_i|\Gamma_i|.\ee

For a 1PI graph $\Gamma$ we let then its superficial degree of
divergence $w(\Gamma)$ be \be
w(\Gamma)=-4|\Gamma|+\sum_{p\in\Gamma^{[1]}_{\rm
int}\cup\Gamma^{[0]}}w(p).\ee Note that all 1PI graphs which have
sdd $\leq 0$ have residue in the above finite set $R$ - we are
dealing with a renormalizable theory. Here we are mainly interested
in the structure of superficially divergent graphs, and hence do not
discuss graphs and Green functions which are superficially
convergent. For all $r\in R$, we let $M_r$ be the set of graphs such
that $ {\bf res}(\Gamma)=r$.

\subsection{Isotopy classes of graphs}
The symmetry factor of a graph $\Gamma$, ${\bf sym}(\Gamma)$, is
defined as usual as the rank of the automorphism group of $\Gamma$.

We consider graphs up to the usual isotopy, for fixed external legs:
\be \qcdbggq=\qcdbggr\not=\qcdbggs=\qcdbggt.\ee This plays a role in
the study of the pre-Lie structure on graphs below. Indeed, we note
that the symmetry factor of a sum of all graphs belonging to an
isotopy class is the product of the symmetry factor of the subgraphs
times the symmetry factor of the cograph obtained by shrinking the
subgraphs. This ensures compatibility of symmetry factors under
graph insertions as in the following example. \be {\bf
sym}\left(\qcdbggq+\qcdbggr\right)=4=\overbrace{{\bf
sym}\left(\qcdagggd\right)}^2\;\times\;\overbrace{{\bf
sym}\left(\qcdaggb\right)}^2,\ee so that \be \frac{1}{4}\left(
\qcdbggq+\qcdbggr\right)=\frac{1}{{\bf
sym}\left(\qcdbggq\right)}\qcdbggq,\ee with \be {\bf
sym}\left(\qcdbggq\right)={\bf sym}\left(\qcdbggr\right)=2.\ee

\subsection{Combinatorial Green functions}
We can now speak of the set of superficially divergent 1PI graphs
and consider graphs according to their residue and loop number
$|\Gamma|$.

We define the formal sums \be \Gamma^{r}=1+\sum_{\Gamma\in
M_r}g^{2\mid\Gamma\mid}\frac{\Gamma}{{\bf sym}(\Gamma)},\;r\in
R_V,\ee \be \Gamma^{r}=1-\sum_{\Gamma\in
M_r}g^{2\mid\Gamma\mid}\frac{\Gamma}{{\bf sym}(\Gamma)},\;r\in
R_E.\label{eq:msign}\ee

We let \be c_{k}^{r}=\sum_{\mid \Gamma\mid=k\atop {\rm
res}(\Gamma)=r}\frac{\Gamma}{{\bf sym}(\Gamma)},\;r\in R,\ee be the
sum of graphs with given residue $r\in R$ and loop number $k$. We
have $\Gamma^{r}=1+\sum_{k=1}^\infty g^{2k}c_k,\;r\in R_V $ and
$\Gamma^{r}=1-\sum_{k=1}^\infty g^{2k}c_k,\;r\in R_E$.

We call these formal sums $\Gamma^{r}$ combinatorial Green
functions. Each term on their rhs maps under the Feynman rules to a
contribution in the perturbative expansion of the Green functions of
our gauge theory. Already the algebraic structure of these
combinatorial Green functions is rather interesting. Analytic
consequences will be briefly discussed at the end and explored in
subsequent work.

Our task in this paper is to acquaint the reader with the structure
of these sums, which is amazingly rich even at this elementary
combinatorial level.
\subsection{Insertion places}
Each graph $\Gamma$ has internal edges $\in \Gamma^{[1]}_{\rm int}$
and vertices $\in \Gamma^{[0]}$. We call (subsets of) those edges and vertices
places of $\Gamma$. Note that each place provides adjacent edges:
for a vertex these are the edges attached to it, while each interior
point of an edge defines two edges adjacent edges for it: the two
pieces of the edge on both sides of that point. In such places,
other graphs can be inserted, using a bijection between the external
edges of those graphs and the adjacent edges provided by these very
places.

The first thing we need to do is to count the number of insertion
places with respect to the graphs to be inserted. Let $X=\prod_i
\Gamma_i$ be a disjoint union of graphs to be inserted. Let us
introduce variables $a_r$ for all $r\in R$. To $X$ we assign the
monomial \be x:=\prod_i a_{{\rm res}(\Gamma_i)}.\ee This monomial
defines integers $n_{X,s}$ for all $s\in R$ by setting \be
x=\prod_{s\in R} a_s^{n_{X,s}}.\ee For example, if
$X=\qcdaggb\qcdaggb$, then \be n_{X,\qcdzggb}=2,\;n_{X,s}=0,\,s\not=
\qcdzggb.\ee Furthermore, to a graph $\Gamma$ assign the function
$b_\Gamma$, and integers $m_{\Gamma,s}$ by \be b_\Gamma:=\prod_{v\in
\Gamma^{[0]}}a_v\prod_{e\in\Gamma^{[1]}_{\rm
int}}\frac{1}{1-a_e}=\prod_{s\in R_V}a_s^{m_{\Gamma,s}}\prod_{e\in
R_E}\frac{1}{[1-a_e]^{m_{\Gamma,e}}}.\ee Then, we define the number
of insertion places for $X$ in $\Gamma$, denoted $\Gamma|X$, by  \be
\Gamma|X:=\prod_{s\in R_V} \left( {m_{\Gamma,s}\atop n_{X,s}}\right)
\prod_{e\in
R_E}\frac{\partial^{n_{X,e}}\frac{1}{[1-a_e]^{m_{\Gamma,e}}}(0)}{n_{X,e}!}.\ee
A few examples: \be\qcdagggc|\qcdaggb=3,
\qcdaggggj|\qcdaggggj=2,\;\qcdaggggj|\qcdaggggj\qcdaggggj=1,\ee\be
\qcdaggb|\qcdaggb\!\qcdaggb=3,\ee as, for the last case, \be
\partial^2_{a_{\qcdzggb}}\frac{1}{[1-a_{\qcdzggb}]^2}(0)=6.\ee

\subsection{Permutation of external edges}
We call $|\Gamma|_\vee$ the number of distinct graphs $\Gamma$ which
are equal upon removal of the external edges. For example \be
\left|\qcdaggggj\!\right|_\vee=\left|\qcdaggggk\!\right|_\vee=\left|\qcdaggggl\!\right|_\vee=3.\ee
Such graphs can be obtained from each other by a permutation of
external edges.

Furthermore for graphs $\gamma_1, \gamma_2, \gamma$ we let ${\bf
top}(\gamma_1,\gamma_2,\gamma)_p$ be the number of bijections
between $\gamma^{[1]}_{2,\rm ext}$ and a chosen place $p={\bf
res}(\gamma_2)\in \gamma_1^{[0]}\cup \gamma_{1,{\rm int}}^{[1]}$
such that $\gamma$ is obtained. This counts the number of ways to
glue $\gamma_2$ into a chosen place $\in \gamma_1$ to obtain
$\gamma$. This has a straightforward generalization ${\bf
top}(\gamma_1,X,\gamma)_p$ to products of graphs $X$, where now $p$
is an appropriate set of chosen residues $\in
\gamma_1^{[0]}\cup\gamma_{1,{\rm int}}^{[1]}$.

For example if $\gamma=\qcdbgga$, $\gamma_1=\qcdaggc$, and $p$ the
vertex $\qcdzggd$ in $\gamma_1$, we have \be {\bf
top}\left(\qcdaggc,\qcdagggga,\qcdbgga\right)_p=2.\ee By definition,
at a given place $p$, \be {\bf top}(\gamma_1,X,\gamma)_p={\bf
top}(\gamma_1,\tilde{X},\gamma)_p,\label{forgins}\ee for all pairs
$X,\tilde{X}$ related by a permutation of external legs.

\subsection{Ramification in graphs}
Above, we have counted the number of ways ${\bf
top}(\gamma_1,X,\gamma)_p$ how to glue graph(s) $X$ into a chosen
single place $p\subset \gamma_1^{[0]}\cup\gamma_{1,{\rm int}}^{[1]}$
so as to obtain a given graph $\gamma$. Furthermore, there might
be various different places $p_i\in \gamma_1$ which provide a
bijection for $X$ such that the same $\gamma$ is obtained.

Let ${\bf bij}(\gamma_1,\gamma_2,\gamma)$ be the number of
bijections between $\gamma_{2,{\rm ext}}^{[1]}$ and adjacent edges
of places $p\sim{\bf res}(\gamma_2)$ in $\gamma_1$ such that
$\gamma$ is obtained.

A graph is described by vertices, edges and relations. For any
bijection as above, we understand that the relations in
$\gamma_2$, together with the relations in $\gamma_1$ which remain
after removal of a chosen place, and the relations provided by the
bijection combine to the relations describing the graph $\gamma$.

We let $\{{\bf bij}\}(\gamma_1,\gamma_2,\gamma)$ be the set of all such
bijections which allow to form $\gamma$ from $\gamma_1$ and $\gamma_2$
and write, for each $b\in \{{\bf
bij}\}(\gamma_1,\gamma_2,\gamma)$, \be \gamma=\gamma_1 b\gamma_2.\ee
We declare ${\bf top}(\gamma_1,\gamma_2,\gamma)_p$ to be the number
of such bijections restricted to a place $p$ in $\gamma_1$.

We have a factorization into the bijections at a given place $p$,
and the distinct bijections which lead to the same result at that
place: \be {\bf bij}(\gamma_1,\gamma_2,\gamma)={\bf
top}(\gamma_1,\gamma_2,\gamma){\bf
ram}(\gamma_1,\gamma_2,\gamma).\label{facttop}\ee Here, ${\bf
ram}(\gamma_1,\gamma_2,\gamma)$ counts the numbers of different
places $p\in \gamma_1^{[0]}\cup\gamma_{1,{\rm int}}^{[1]}$ which
allow for bijections such that \be \gamma_1 b \gamma_2=\gamma.\ee

Note that for any two such places $p,\tilde{p}$ we find precisely
${\bf top}(\gamma_1,\gamma_2,\gamma)$ such bijections: \be {\bf
top}(\gamma_1,\gamma_2,\gamma):={\bf
top}(\gamma_1,\gamma_2,\gamma)_p={\bf
top}(\gamma_1,\gamma_2,\gamma)_{\tilde{p}}.\ee

 One
immediately confirms that this number is indeed independent of  the place $p$
as we can pair off the bijections at $p$ with the bijections at
$\tilde{p}$  for any places $p,p^\prime$, so that the
factorization (\ref{facttop}) of ${\bf
bij}(\gamma_1,\gamma_2,\gamma)$ is straightforward.

We call this integer ${\bf ram}(\gamma_1,\gamma_2;\gamma)$ the
ramification index: it counts the degeneracy of inserting a graph
at different places - if the ramification index is greater than
one, the same graph $\Gamma$ can be obtained from inserting a
graph $\gamma_2$ into a graph $\gamma_1$ at different places. For
example \be {\bf
ram}\left(\qcdaggb,\qcdagggc,\qcdbggg\right)=2,\;{\bf
ram}\left(\qcdaggc,\qcdagggga,\qcdbggg\right)=1 .\ee The
generalization replacing $\gamma_2$ by a product of graphs $X$ is
straightforward. The motivation of the name comes from a
comparison with the situation in the study of number fields which
will be given in future work.

\subsection{pre-Lie structure of graphs}
The pre-Lie product we will use is a sum over all bijections and
places of graph insertions. Hence it gives the same result for the
insertion of any two graphs related by permutation of their external
legs. One could formulate the Hopf and Lie structure hence on graphs
with amputated external legs, but we will stick with the usual
physicists convention and work with Feynman graphs which have
external edges.

We define $n(\gamma_1,X,\Gamma)$ as the number of ways to shrink $X$
to its residue (a set of one or more places) in $\Gamma$ such that
$\gamma_1$ remains.

 We  define a bilinear map \be
\Gamma_1*\Gamma_2=\sum_\Gamma \frac{n(\Gamma_1,\Gamma_2,\Gamma)
}{|\Gamma_2|_\vee}\Gamma. \ee This is a finite sum, as on the rhs
only graphs can contribute such that \be
|\Gamma|=|\Gamma_1|+|\Gamma_2|.\ee We divide by the number of
permutations of external edges  $|\Gamma_2|_\vee$ to eliminate the
degeneracy in $n(\Gamma_1,\Gamma_2,\Gamma)$, a number which is
insensitive to the orientation of edges of $\Gamma_2$. Note that for
$\Gamma_a \sim_{\rm perm}\Gamma_b$, we have $\Gamma_1 * \Gamma_a=
\Gamma_1 * \Gamma_b$. Here, $\sim_{\rm perm}$ indicates equivalence
upon permutation of external edges.

For example, \be \qcdaggb*\qcdagggc=2\qcdbggg.\ee while \be
\qcdaggb*(\qcdaggga+\qcdagggb)=\qcdbggh+\qcdbggi+\qcdbggj+\qcdbggk.\ee
\begin{prop}

This map is pre-Lie: \be
(\Gamma_1*\Gamma_2)*\Gamma_3-\Gamma_1*(\Gamma_2*\Gamma_3)=
(\Gamma_1*\Gamma_3)*\Gamma_2-\Gamma_1*(\Gamma_3*\Gamma_2). \ee
\end{prop}
Note that the graphs on the rhs have all the same residue as
$\Gamma_1$. The proof is analogous to the one in
\cite{Connes:1999yr}. For a product of graphs $X$ we define
similarly \be \Gamma_1*X=\sum_\Gamma
\frac{n(\Gamma_1,X,\Gamma)}{|X|_\vee}\Gamma. \ee This is a
straightforward generalization of this map, but certainly not a
pre-Lie product in that generality.
\subsection{The Lie algebra of graphs}

We let ${\cal L}$ be the corresponding Lie algebra, obtained by
antisymmetrizing the  pre-Lie product: \be
[\Gamma_1,\Gamma_2]=\Gamma_2*\Gamma_1-\Gamma_1*\Gamma_2.\ee The
bracket $[,]$ fulfils a Jacobi identity and we hence get a graded
Lie algebra. Note that the terms generated by the Lie bracket
involve graphs of different residue.

\subsection{The Hopf algebra of graphs}
Let  ${\cal H}$ be the corresponding Hopf algebra. Let us quickly
describe how it is found.
 To ${\cal L}$, we assign its universal
enveloping algebra \be U({\cal L})=\bigoplus_{j=0}^\infty T({\cal
L})^{(j)},\ee where $T({\cal L})^{(j)}={\cal L}^{\otimes j}$ is
the $j$-fold tensorproduct of ${\cal L}$. In $U({\cal L})$ we
identify \be
\Gamma_1\otimes\Gamma_2-\Gamma_2\otimes\Gamma_1=[\Gamma_1,\Gamma_2],\ee
as usual. We let \be \langle \Gamma_1,\Gamma_2\rangle=\left\{
{0,\;\Gamma_1\not=\Gamma_2} \atop {1,\; \Gamma_1=\Gamma_2} \right.
. \ee Here we understand that entries on the lhs of
$\langle\cdot,\cdot\rangle$ belong to the Lie algebra, entries on
the rhs to the Hopf algebra.

We compute the coproduct from this pairing requiring \be \langle
[\Gamma_1,\Gamma_2],\Delta(\Gamma)\rangle=\langle
\Gamma_1\otimes\Gamma_2-\Gamma_2\otimes\Gamma_1,\Delta(\Gamma)\rangle,\ee
and find the usual composition into subgraphs and cographs \be
\Delta(\Gamma)=\Gamma\otimes
1+1\otimes\Gamma+\sum_\gamma\gamma\otimes \Gamma/\gamma.\ee The
antipode $S:{\cal H}\to {\cal H}$ is \be
S(\Gamma)=-\Gamma-\sum_{\gamma}S(\gamma) \Gamma/\gamma.\ee The
counit $\bar{e}$ annihilates the augmentation ideal as usual
\cite{Connes:1999yr,Kreimer:2002rf}.

Furthermore, we define $|\Gamma|_{\rm aug}$ to be the augmentation
degree, defined via the projection $P$ into the augmentation ideal.
Furthermore, for future use we let $c_{k,s}^{r}$ be the sum of graphs with given residue $r$,
loop number $k$ and augmentation degree $s$. ${\cal H}_{\rm lin}$ is
the span of the linear generators of ${\cal H}$.

With the Hopf algebra comes its character group, and with it three
distinguished objects: the Feynman rules $\phi$, the $\bar{R}$
operation \be \bar{\phi}=m(S_R^\phi\otimes\phi
P)\Delta,\label{eq:rbar}\ee (which is a character with regard to the
double structure of Rota--Baxter algebras \cite{FGK}) and
counterterm $-R\bar{\phi}$.

Note that the forgetfulness upon insertion wrt the external legs
(\ref{forgins}) forces us to work with a symmetric renormalization
scheme \be S_R^\phi(\Gamma_1)=S_R^\phi(\Gamma_2),\ee for
consistency, for all pairs $\Gamma_1,\Gamma_2$ which agree by a
permutation of external edges.

Indeed, $\forall \gamma$ and $\Gamma_1\sim_{\rm perm}\Gamma_2$,
\bea 0 & = & \gamma*\Gamma_1-\gamma*\Gamma_2\\
 & = & \bar{\phi}(\gamma*\Gamma_1-\gamma*\Gamma_2)\\
  & = & S_R^\phi(\Gamma_1-\Gamma_2)\phi(\gamma),
\eea upon using (\ref{eq:rbar}) and as
$\phi(\gamma*\Gamma_1-\gamma*\Gamma_2)=0$, and similarly for all
cographs of $\Gamma_1$ and $\Gamma_2$.

\subsection{External structures}
In later work it will be useful to disentangle Green functions wrt
to their form-factor decomposition. This can be easily achieved by
the appropriate use of external structures \cite{Connes:1999yr}.

We hence extend graphs $\gamma$ to pairs $(\gamma,\sigma)$ where
$\sigma$ labels the formfactor and with a forgetful rule \be
\sum_{\sigma_2}(\Gamma_1,\sigma_1)*(\Gamma_2,\sigma_2):=\sum_\Gamma
\frac{n(\Gamma_1,\Gamma_2,\Gamma)}{|\Gamma_2|_\vee}(\Gamma,\sigma_1).
\ee This allows to separate the form-factor decompositions as
partitions of unity  $1=\sum_{\sigma_2}$ in computationally
convenient ways for which we will use in future work. If we do not
sum over $\sigma_2$ we can extend our notation to marked graphs as
in \cite{Connes:1999yr}.
\section{The theorems} In this section we state
the main result. It concerns the role played by the maps $B_+^{k;r}$
to be defined here: they provide the equations of motion, ensure
locality, and lead us to the Slavnov--Taylor identities for the
couplings.

We start by defining a map \be B_+^{k;r}=\sum_{{\mid\gamma\mid=k
\atop\mid\gamma\mid_{\rm aug}=1} \atop{\rm
res}(\gamma)=r}\frac{1}{{\bf sym}(\gamma)}B_+^\gamma,\ee where
$B_+^\gamma$ is a normalized generalization of the pre-Lie insertion
into $\gamma$ defined by requiring $B_+^{k;r}$ to be Hochschild
closed. To achieve this, we need to count the maximal forests of a
graph $\Gamma$. It is the number of ways to shrink subdivergences to
points such that the resulting cograph is primitive. To define it
more formally, we use Sweedler's notation to write $\Delta(X)=\sum
X^\prime\otimes X^{\prime\prime}$. If $X=\prod \Gamma_i$ is a Hopf
algebra element with $\Gamma_i$ graphs we write \be
\Delta(X)=c(X^\prime,X^{\prime\prime}) \widehat{X^\prime}\otimes
\widehat{X^{\prime\prime}},\ee which defines scalars
$c(X^\prime,X^{\prime\prime})$. Here, $\widehat{X^\prime}$ and
$\widehat{X^{\prime\prime}}$ are graphs and the section coefficients
of the Hopf algebra $c(X^\prime,X^{\prime,\prime})$ are explicitely
spelled out.

We now set
 \be {\rm
maxf}(\Gamma)=\sum_{|\gamma|_{\rm aug}=1} \sum
c(\Gamma^\prime,\Gamma^{\prime\prime})\langle\gamma,\Gamma^{\prime\prime}\rangle.\ee
Note that this counts precisely the ways of shrinking subgraphs to
points such that a primitive cograph remains, as it should, using
the pairing between the Lie and Hopf algebra and summing over all
Lie algebra generators indexed by primitive graphs $\gamma$.

The same number can by definition be obtained from the section
coefficients of the pre-Lie algebra: \be {\rm
maxf}(\Gamma)=\sum_{|\gamma|_{\rm
aug}=1}\sum_{|X|=|\Gamma|-|\gamma|}n(\gamma,X,\Gamma),\ee as each
maximal forest has a primitive cograph $\gamma$ and some
subdivergences $X$ of loop number $|\Gamma|-|\gamma|$.

We have defined the pre-Lie product so that  \be
\gamma*X=\sum_{\Gamma\in {\cal H}_{\rm
lin}}\frac{n(\gamma,X,\Gamma)}{|X|_\vee}\Gamma.\ee Now we define \be
B_+^\gamma(X)=\sum_{\Gamma\in {\cal H}_{\rm lin}}\frac{{\bf bij
}(\gamma,X,\Gamma)}{|X|_\vee}\frac{1}{{\rm
maxf}(\Gamma)}\frac{1}{\left[ \gamma|X\right]}\Gamma,\ee for all $X$
in the augmentation ideal. Furthermore, $B_+^\gamma(\One)=\gamma$.

Taking into account the fact that the pre-Lie product is a sum over
all labelled composition of graphs and the fact that we carefully
divide out the number of possibilities to generate the same graph,
we can apply the corresponding results for rooted trees
\cite{Bergbauer:2005fb}. One concludes in analogy to Theorem
\ref{thm:dsesub}:
\begin{theorem} (the Hochschild theorem)\\
\be \Gamma^{r}\equiv 1+\sum_{\Gamma\in M_r}\frac{\Gamma}{{\bf
sym}(\Gamma)}=1+\sum_{k=1}^\infty g^k \sum_{{\mid\gamma\mid=k
\atop\mid\gamma\mid_{\rm aug}=1} \atop{\rm
res}(\gamma)=r}\frac{1}{{\bf sym}(\gamma)}B_+^\gamma (X_\gamma),\ee
where $X_\gamma=\prod_{v\in\gamma^{[0]}}\Gamma^v\prod_{e\in
\gamma_{\rm int}^{[1]}}1/\Gamma^e$.
\end{theorem}

 For the
next theorem, we have to define the Slavnov--Taylor identities for
the coupling. Consider \be X_{k,r}=\Gamma^{r}X_{\rm coupl}^k.\ee We
set \be X_{\rm coupl}=1+\sum_{k=1}^\infty g^{2k} c^{\rm coupl}_k,\ee
which determines the $c^{\rm coupl}_k$ as polynomials in the $c^r_j$
from the definition of $X_{\rm coupl}$ below. The Slavnov--Taylor
identities for the couplings can be written as \be
\frac{\Gamma^{\qcdzgff}}{\Gamma^{\qcdzff}}=\frac{\Gamma^{\qcdzguu}}{\Gamma^{\qcdzuu}}
=\frac{\Gamma^{\qcdzggc}}{\Gamma^{\qcdzggb}}=\frac{\Gamma^{\qcdzggd}}{\Gamma^{\qcdzggc}},\label{ST}\ee
which results in identities in every order in $g^2$ and leads to
define indeed a single coupling $X_{\rm coupl}$ which can be defined
in four equal ways, each one corresponding to an interaction
monomial in the Lagrangian: \be X_{\rm
coupl}=\frac{\Gamma^{\qcdzgff}}{\Gamma^{\qcdzff}\sqrt{\Gamma^{\qcdzggb}}}=\frac{\Gamma^{\qcdzguu}}{\Gamma^{\qcdzuu}\sqrt{\Gamma^{\qcdzggb}}}
=\frac{\Gamma^{\qcdzggc}}{[\Gamma^{\qcdzggb}]^{3/2}}=\frac{\sqrt{\Gamma^{\qcdzggd}}}{\Gamma^{\qcdzggb}}.\ee
Note that we read this identities as describing the kernel of the
counterterm: they hold under the evaluation of the indicated series
of graphs by the corresponding character $S_R^\phi$. The following
theorem follows on imposing these identities as relations between
Hopf algebra elements order by order in $g^2$. On the other hand, if
we assume the following theorem, we would derive the existence of
the Slavnov--Taylor identities from the requirement of the existence
of the grading sub Hopf algebra furnished by the elements $c_k^r$.

\begin{theorem} (the gauge theory theorem)\\
\be i)\; \Gamma^{r}\equiv 1+\sum_{\Gamma\in M_r}\frac{\Gamma}{{\bf
sym}(\Gamma)}=1+\sum_{k=1}^\infty g^k B_+^{k;r}(X_{k,r})\ee \be
ii)\; \Delta(B_+^{k;r}(X_{k,r}))=\One\otimes
B_+^{k;r}(X_{k,r})+({\rm id}\otimes B_+^{k;r})\Delta(X_{k,r}).\ee
\be iii)\; \Delta(c^{r}_{k})=\sum_{j=0}^k {\rm Pol}^{r}_j(c)\otimes
c^{r}_{k-j},\ee where ${\rm Pol}^{r}_j(c)$ is a polynomial in the
variables $c_m^r$ of degree $j$, determined as the order $j$
coefficient in the Taylor expansion of $\Gamma^r [X_{\rm coupl}]^j$.
\end{theorem}
\end{fmffile}
\begin{fmffile}{fmfG-Exa1}
\end{fmffile}

\section{Two-loop Example}
\subsection{One-loop graphs}
\begin{fmffile}{fmfG-Exa2}
This section provides an instructive example. We consider our
non-abelian gauge theory and first list its one-loop graphs, which
provide by definition maps from ${\cal H}\to {\cal H}_{\rm lin}$.

The maps $B_+^{k,r}$, we claim, furnish the Hochschild one-cocyles
and provide the Dyson--Schwinger equations, in accordance with the
Hochschild and gauge theorems. We study this for the self-energy of
the gauge boson to two-loops. We want in particular exhibit the fact
that each such two-loop graph is a sum of terms each lying in the
image of such a map and want to understand the role of Hochschild
cohomology.

We have for example \be
B_+^{1,\qcdzggb}=\frac{1}{2}B_+^{\qcdaggb}+\frac{1}{2}B_+^{\qcdaggc}+B_+^{\qcdagga}+B_+^{\qcdaggd}.\ee

To find the one-loop graphs we simply have to apply these maps to
the unit of the Hopf algebra of graphs, which is trivial: \bea
c_1^{\qcdzggb} & = &  B_+^{1,\qcdzggb}(\One) = B_+^{\qcdagga}(\One)+\frac{1}{2}B_+^{\qcdaggb}(\One)+\frac{1}{2}B_+^{\qcdaggc}(\One)+B_+^{\qcdaggd}(\One)\nonumber\\
& = & \qcdagga+\frac{1}{2}\qcdaggb+\frac{1}{2}\qcdaggc+\qcdaggd\eea
and similarly \bea
c_1^{\qcdzff} & = & B_+^{\qcdaffa}\!(\One)=\qcdaffa\\
c_1^{\qcdzuu} & = & B_+^{\qcdauua}\!(\One)=\qcdauua\\
c_1^{\qcdzggc}& = &
B_+^{\qcdaggga}(\One)+B_+^{\qcdagggb}(\One)+B_+^{\qcdagggc}(\One)\nonumber\\
& + & \frac{1}{2}\left[B_+^{\qcdagggd}(\One) + B_+^{\qcdaggge}(\One)+B_+^{\qcdagggf}(\One)\right]\nonumber\\ & + &
B_+^{\qcdagggg}(\One)+B_+^{\qcdagggh}(\One)\nonumber\\
& = & \qcdaggga+\qcdagggb+\qcdagggc+\frac{1}{2}\left[\qcdagggd\right.\nonumber\\ & + & \left. \qcdaggge+\qcdagggf\right]+\qcdagggg+\qcdagggh\\
c_1^{\qcdzggd}& = & B_+^{\qcdagggga}(\One)+B_+^{\qcdaggggb}(\One)+B_+^{\qcdaggggc}(\One)+B_+^{\qcdaggggd}(\One)\nonumber\\
& + &
 B_+^{\qcdagggge}(\One)+B_+^{\qcdaggggf}(\One)+B_+^{\qcdaggggg}(\One)+B_+^{\qcdaggggh}(\One)\nonumber\\
& + &
B_+^{\qcdaggggi}(\One)+\frac{1}{2}\left[B_+^{\qcdaggggj}(\One)+B_+^{\qcdaggggk}(\One)+B_+^{\qcdaggggl}(\One)\right]\nonumber\\
& + &
B_+^{\qcdaggggm}(\One)+B_+^{\qcdaggggn}(\One)+B_+^{\qcdaggggo}(\One)+B_+^{\qcdaggggp}(\One)\nonumber\\
& + &
B_+^{\qcdaggggq}(\One)+B_+^{\qcdaggggr}(\One)+B_+^{\qcdaggggs}(\One)+B_+^{\qcdaggggt}(\One)\nonumber\\
& + &
B_+^{\qcdaggggu}(\One)+B_+^{\qcdaggggv}(\One)+B_+^{\qcdaggggw}(\One)+B_+^{\qcdaggggx}(\One)\nonumber\\
& = &
\qcdagggga+\qcdaggggb+\qcdaggggc+\qcdaggggd\nonumber\\
& + &
\qcdagggge+\qcdaggggf+\qcdaggggg+\qcdaggggh\nonumber\\
& + &
\qcdaggggi+\frac{1}{2}\left[\qcdaggggj+\qcdaggggk+\qcdaggggl\right]\nonumber\\
& + &
\qcdaggggm+\qcdaggggn+\qcdaggggo+\qcdaggggp\nonumber\\
& + &
\qcdaggggq+\qcdaggggr+\qcdaggggs+\qcdaggggt\nonumber\\
& + & \qcdaggggu+\qcdaggggv+\qcdaggggw+\qcdaggggx. \eea
\end{fmffile}
\begin{fmffile}{fmfG-Exa3}
\subsection{Two-loop graphs}
We now want to calculate \bea c_2^{\qcdzggb} & = &
B_+^{\qcdagga}(2c_1^{\qcdzgff}+2c_1^{\qcdzff})+
\frac{1}{2}B_+^{\qcdaggb}(2c_1^{\qcdzggc}+2c_1^{\qcdzggb})\nonumber\\
& + &
\frac{1}{2}B_+^{\qcdaggc}(c_1^{\qcdzggd}+c_1^{\qcdzggb})+B_+^{\qcdaggd}(2c_1^{\qcdzguu}+2c_1^{\qcdzuu}),\eea
upon expanding \bea X_{\qcdaggb} & = &
\frac{\left[\Gamma^{\qcdzggc}\right]^2}{\left[\Gamma^{\qcdzggb}\right]^2},\\
X_{\qcdaggc} & = & \frac{\Gamma^{\qcdzggd}}{\Gamma^{\qcdzggb}},\\
X_{\qcdaggd} & = & \frac{\left[\Gamma^{\qcdzguu}\right]^2}{\left[\Gamma^{\qcdzuu}\right]^2},\\
X_{\qcdagga} & = &
\frac{\left[\Gamma^{\qcdzgff}\right]^2}{\left[\Gamma^{\qcdzff}\right]^2},
\eea to order $g^2$.

Let us do this step by step. Adding up the contributions, we should
find precisely the two-loop contributions to the gauge-boson
self-energy, and the coproduct \be \Delta(c_2^{\qcdzggb})  =
c_2^{\qcdzggb}\!\otimes\One+\One\otimes c_2^{\qcdzggb}\!+[2c_1^{\rm
coupl}-c_1^{\qcdzggb}]\otimes c_1^{\qcdzggb}.\ee The minus sign
appears on the rhs due to our conventions in (\ref{eq:msign}).\\
{\bf Insertions in $\frac{1}{2} B_+^{\qcdaggb}$\\}
\end{fmffile}
Below, we will give coefficients like
$\left(\frac{1}{2}|1|2|\frac{1}{2}|\frac{1}{2}|1|1\right)$ in the
next equation, where the first entry is the symmetry factor of the
superscript $\gamma$ of $B_+^\gamma$, the second entry the symmetry
factor of the graphs in the argument $X$, the third entry the
integer weight of that argument, the fourth entry the number of
insertion places, the fifth entry the number of maximal forests of
the graphs $\Gamma$ on the rhs, the sixth entry is ${\bf
top}(\gamma,X,\Gamma)$ and the seventh entry is ${\bf
ram}(\gamma,X,\Gamma)$.
\begin{fmffile}{fmfG-Exa4}

We start \bea
\frac{1}{2}B_+^{\qcdaggb}\left(2\qcdaggga+2\qcdagggb\right) & =
& \left(\frac{1}{2}|1|2|\frac{1}{2}|\frac{1}{2}|1|1\right)\nonumber\\
& \times & \left(
\qcdbggh+\qcdbggi+\qcdbggj+\qcdbggk\right)\nonumber\\
 & = & \frac{1}{4}\left(
\qcdbggh+\qcdbggi+\qcdbggj+\qcdbggk\right)\nonumber\\
 & = & \frac{1}{2}\left(\qcdbggh + \qcdbggj \right),\label{97}\eea where indeed the
symmetry factor for $\qcdaggb$ is $1/2$, the symmetry factor for the
graphs appearing as argument is 1, and they appear with weight two.
We have two three-gluon vertices in $\qcdaggb$, and hence two
insertion places. Each graph on the right has two maximal forests,
and for each graph the inserted subgraph can be reduced in a unique
way to obtain $\qcdaggb$, so the ramification index is one, and the
topological weight is unity as well.

Similarly for ghosts \bea
\frac{1}{2}B_+^{\qcdaggb}\left(2\qcdagggg+2\qcdagggh\right) & = &
\left(\frac{1}{2}|1|2|\frac{1}{2}|\frac{1}{2}|1|1\right)\nonumber\\
& \times & \left(
\qcdbggl+\qcdbggm+\qcdbggn+\qcdbggo\right)\nonumber\\
 & = & \frac{1}{4}\left(
\qcdbggl+\qcdbggm+\qcdbggn+\qcdbggo\right)\nonumber\\
 & = & \frac{1}{2}\left(\qcdbggl + \qcdbggn \right).\label{98}\eea

Next, \bea \frac{1}{2}B_+^{\qcdaggb}\left(2\qcdagggc\right) & = &
\left(\frac{1}{2}|1|2|\frac{1}{2}|\frac{1}{3}|1|2\right)
\qcdbggg\nonumber\\
 & = & \frac{1}{3}
\qcdbggg.\label{99}\eea This is more interesting. There are three maximal
forests in $\qcdbggg$, a one-loop three-point vertex-graph to the
left and to the right, and also the four gluon propagators which
form the square give a one-loop log-divergent four-point graph.
Also, we can form $\qcdbggg$ by inserting the argument into either
vertex of $\qcdaggb$, and hence the ramification index is two. The
topological index is 1. Note that the total weight $1/3$ of the
graph is not its contribution to $c_2^{\qcdzggb}$. We expect the
same graph to be generated from inserting into $\qcdaggc$, as we
will confirm soon. This is generally true: only in the Hochschild
closed sum over insertions in all components of $c_1^{\qcdzggb}$
will we see the correct weights emerging.

We continue. \bea
\frac{1}{2}B_+^{\qcdaggb}\left(2\left(\frac{1}{2}\qcdagggd+\frac{1}{2}\qcdaggge+\frac{1}{2}\qcdagggf\right)\right)
& = &
\nonumber\\
\left(\frac{1}{2}|\frac{1}{2}|2|\frac{1}{2}|\frac{1}{2}|2|1\right)
\qcdbggp & &\nonumber\\
  +  \left(\frac{1}{2}|\frac{1}{2}|2|\frac{1}{2}|\frac{1}{3}|1|2\right)
\left( \qcdbggq+\qcdbggs\right)
& & \nonumber\\
  =  \frac{1}{4}\left(
\qcdbggp\right)+\frac{1}{6}\left( \qcdbggq+\qcdbggs\right). &
& \label{100}\eea Here, the first graph $\qcdbggp$ on the rhs has two maximal
forests, a ramification index of two as the graph can be obtained
by insertion in either vertex of $\qcdaggb$, and $\qcdbggp$ is
generated by one bijection, while the other two graphs have three
maximal forests, 1 as a ramification index and 2 as a topological
index: there are two different bijections leading to each of them.

So far we inserted 3-point one-loop vertex corrections. Now we
insert propagator corrections.

\bea
\frac{1}{2}B_+^{\qcdaggb}\left(2\left(\frac{1}{2}\qcdaggb\right)\right)
& = &
\left(\frac{1}{2}|\frac{1}{2}|2|\frac{1}{2}|\frac{1}{3}|1|1\right)\left(
\qcdbgga+\qcdbggb\right)\nonumber\\
 & = & \frac{1}{12}\left(
\qcdbgga+\qcdbggb\right)=\frac{1}{6}\qcdbgga.\label{101}\eea Note that the
graph allows for three maximal forests: apart from the inserted
one-loop self-energy graph it has two more maximal forests,
corresponding to the two one-loop four-point vertex-subgraphs in $
\qcdbgga$, obtained by opening an internal edge in the subgraph.

Next we insert a fermion loop:
 \bea \frac{1}{2}B_+^{\qcdaggb}\left(2\qcdagga\right) & = &
\left(\frac{1}{2}|1|2|\frac{1}{2}|1|1|1\right)\left(
\qcdbggc+\qcdbggd\right)\nonumber\\
 & = & \frac{1}{2}\left(
\qcdbggc+\qcdbggd\right)=\qcdbggc.\label{102}\eea Indeed, no ramification, just
one maximal forest and a single bijection. Similarly, the ghost loop
\bea \frac{1}{2}B_+^{\qcdaggb}\left(2\qcdaggd\right) & = &
\left(\frac{1}{2}|1|2|\frac{1}{2}|1|1|1\right)\left(
\qcdbgge+\qcdbggf\right)\nonumber\\
 & = & \frac{1}{2}\left(
\qcdbgge+\qcdbggf\right)=\qcdbgge.\label{103}\eea Finally, \bea
\frac{1}{2}B_+^{\qcdaggb}\left(2\left( \frac{1}{2}\qcdaggc\right)
\right) & = &
\left(\frac{1}{2}|\frac{1}{2}|2|\frac{1}{2}|\frac{1}{2}|1|1\right)\left(
\qcdbggaa+\qcdbggab\right)\nonumber\\
& = & \frac{1}{8}\left( \qcdbggaa+\qcdbggab\right)\nonumber\\
 & = & \frac{1}{4}\left(
\qcdbggaa\right).\label{104}\eea A single bijection, no ramification and two
maximal forests in $\qcdbggaa$. This concludes insertions into
$\qcdaggb$.\\[5mm]
\end{fmffile}
\begin{fmffile}{fmfG-exa5}
{\bf Insertions into $\frac{1}{2}B_+^{\qcdaggc}$\\} We come to
insertions into $\qcdaggc$. \bea
\frac{1}{2}B_+^{\qcdaggc}\left(\qcdagggga+\qcdaggggb+\qcdaggggc\right)
& = & \left(\frac{1}{2}|1|1|1|\frac{1}{3}|1|2\right)
\qcdbgga\nonumber\\
 & + & \left(\frac{1}{2}|1|1|1|\frac{1}{3}|1|1\right) \qcdbggg
\nonumber\\
 & = & \frac{1}{3}
\qcdbgga+\frac{1}{6}\qcdbggg.\label{105}\eea Indeed, $\qcdbgga$ has three
maximal forests, no ramification as there is only a single insertion
place and two of the three bijections lead to it, while one
bijection leads to $\qcdbggg$, which also has three maximal forests.
\bea
\frac{1}{2}B_+^{\qcdaggc}\left(\qcdaggggd+\qcdagggge+\qcdaggggf+\qcdaggggg+\qcdaggggh+\qcdaggggi\right)
& = &
\nonumber\\\left(\frac{1}{2}|1|1|1|\frac{1}{3}|1|1\right)\qcdbggac+
\left(\frac{1}{2}|1|1|1|\frac{1}{2}|1|1\right)\qcdbggaa & &\nonumber\\
 +
\left(\frac{1}{2}|1|1|1|\frac{1}{3}|1|2\right)\left(\qcdbggq+\qcdbggs\right) & &\nonumber\\
=\frac{1}{6}\qcdbggac+\frac{1}{4}\qcdbggaa+\frac{1}{3}\left(\qcdbggq+\qcdbggs+\right).
& & \label{106}\eea Note that $\qcdbggac$ has three maximal forests and comes
from one bijection, $\qcdbggaa$ has two maximal forests and comes as
well from one bijection, while each of $\qcdbggq$ and $\qcdbggs$
come from two bijections and have three maximal forests. \bea
\frac{1}{2}B_+^{\qcdaggc}\left(\frac{1}{2}\left(
\qcdaggggj+\qcdaggggk+\qcdaggggl\right)\right) & = &
\nonumber\\\left(\frac{1}{2}|\frac{1}{2}|1|1|\frac{1}{2}|1|1\right)\qcdbggaf+
\left(\frac{1}{2}|\frac{1}{2}|1|1|\frac{1}{3}|1|2\right)\qcdbggag & & \nonumber\\
=\frac{1}{8}\qcdbggaf+\frac{1}{6}\qcdbggag. & & \label{107}\eea This time,
$\qcdbggaf$ has two maximal forests and comes from one bijection
while $\qcdbggag$ has three maximal forests and the two remaining
bijections are leading to it. \bea
\frac{1}{2}B_+^{\qcdaggc}\left(\qcdaggggm+\qcdaggggn+\qcdaggggo+\qcdaggggp+\qcdaggggq+\qcdaggggr\right)
& = &
\nonumber\\\left(\frac{1}{2}|1|1|1|\frac{1}{2}|1|2\right)\left(\qcdbggv+\qcdbggu\right)+
\left(\frac{1}{2}|1|1|1|\frac{1}{3}|1|2\right)\qcdbggw & &\nonumber\\
=\frac{1}{2}\left(\qcdbggv+\qcdbggu\right)+\frac{1}{3}\qcdbggz. &
& \label{108}\eea Indeed, $\qcdbggv$ and $\qcdbggu$ have both two maximal
forests and two bijections leading to them each, while $\qcdbggw$
has three maximal forests and is generated from two bijections.

Similarly for ghosts \bea
\frac{1}{2}B_+^{\qcdaggc}\left(\qcdaggggs+\qcdaggggt+\qcdaggggu+\qcdaggggv+\qcdaggggw+\qcdaggggx\right)
& = &\nonumber\\
\left(\frac{1}{2}|1|1|1|\frac{1}{2}|1|2\right)\left(\qcdbggy+\qcdbggx\right)+
\left(\frac{1}{2}|1|1|1|\frac{1}{3}|1|2\right)\qcdbggw & &\nonumber\\
=\frac{1}{2}\left(\qcdbggy+\qcdbggx\right)+\frac{1}{3}\qcdbggw. &
& \label{109}\eea

Now insertion of self-energies. \bea
\frac{1}{2}B_+^{\qcdaggc}\left(\frac{1}{2}\qcdaggb\right) & = &
\left(\frac{1}{2}|\frac{1}{2}|1|1|\frac{1}{3}|1|1\right)\qcdbggac=\frac{1}{12}\qcdbggac,\label{110}\eea
straightforward.  \bea
\frac{1}{2}B_+^{\qcdaggc}\left(\frac{1}{2}\qcdaggc\right) & = &
\left(\frac{1}{2}|\frac{1}{2}|1|1|\frac{1}{2}|1|1\right)\qcdbggaf=\frac{1}{8}\qcdbggaf,\label{111}\eea
dito. Next, \bea \frac{1}{2}B_+^{\qcdaggc}\left(\qcdagga\right) & =
&
\left(\frac{1}{2}|1|1|1|1|1|1\right)\qcdbggad=\frac{1}{2}\qcdbggad\label{112}\eea
and similar for the ghost-loop \bea
\frac{1}{2}B_+^{\qcdaggc}\left(\frac{1}{2}\qcdaggd\right) & = &
\left(\frac{1}{2}|1|1|1|1|1|1\right)\qcdbggae=\frac{1}{2}\qcdbggae.\label{113}\eea
This concludes insertions into $\qcdaggc$.\\[5mm]
\end{fmffile}
\begin{fmffile}{fmfG-exa6}
{\bf Insertions into $B_+^{\qcdagga}$ and $B_+^{\qcdaggd}$\\} It
remain the insertions into $\qcdagga$ and $\qcdaggd$. \bea
B_+^{\qcdagga}\left(2\qcdagffa\right) & = &
\left(1|1|2|\frac{1}{2}|\frac{1}{3}|2|1\right)\qcdbggz=\frac{2}{3}\qcdbggz.\label{114}\eea
Indeed, there are three maximal forests, a ramification index of two
and just a single bijection for each place.

Next \bea B_+^{\qcdagga}\left(2\qcdagffb\right) & = &
\left(1|1|2|\frac{1}{2}|\frac{1}{2}|1|1\right)\left(\qcdbggh+\qcdbggj\right)\nonumber\\
&  = & \frac{1}{2}\left(\qcdbggh+\qcdbggj\right).\label{115}\eea This time we
have no ramification and two maximal forests.

Next the self-energy, \bea B_+^{\qcdagga}\left(2\qcdaffa)\right) &
= &
\left(1|1|2|\frac{1}{2}|\frac{1}{2}|1|1\right)\left(\qcdbggu+\qcdbggv\right)\nonumber\\
&  = & \frac{1}{2}\left(\qcdbggu+\qcdbggv\right).\label{116}\eea Again, two
maximal forests, single bijection and no ramification.

Finally, the ghosts bring nothing new: \bea
B_+^{\qcdaggd}\left(2\qcdaguua)\right) & = &
\left(1|1|2|\frac{1}{2}|\frac{1}{3}|2|1\right)\qcdbggw=\frac{2}{3}\qcdbggw.\label{117}\eea
And \bea B_+^{\qcdaggd}\left(2\qcdaguub)\right) & = &
\left(1|1|2|\frac{1}{2}|\frac{1}{2}|1|1\right)\left(\qcdbggl+\qcdbggm\right)\nonumber\\
& = & \frac{1}{2}\left(\qcdbggl+\qcdbggm\right).\label{118}\eea Also, \bea
B_+^{\qcdaggd}\left(2\qcdauua)\right) & = &
\left(1|1|2|\frac{1}{2}|\frac{1}{2}|1|1\right)\left(\qcdbggx+\qcdbggy\right)\nonumber\\
&  = & \frac{1}{2}\left(\qcdbggx+\qcdbggy\right).\label{119}\eea
\subsection{Adding up}
 Now we indeed confirm that the results adds up to
$c_2^{\qcdzggb}$. Adding up, we indeed find \beas \frac{1}{2}\left(\qcdbggh
+\qcdbggi+ \qcdbggj+ \qcdbggk \right)
& & {\rm from}\; (\ref{97})+ (\ref{115})\\
+\frac{1}{2}\left(\qcdbggl +\qcdbggm+ \qcdbggn+ \qcdbggo \right)
& &  {\rm from}\;(\ref{98}) + (\ref{118})\\
+\frac{1}{2}\qcdbggg
& &  {\rm from}\;(\ref{99})+(\ref{105})\\
+\frac{1}{4}\qcdbggp
& &  {\rm from}\;(\ref{100})\\
+\frac{1}{2}\left(\qcdbggq+\qcdbggs\right)
& &  {\rm from}\;(\ref{100})+(\ref{106})\\
+\frac{1}{2}\qcdbgga
& &  {\rm from}\;(\ref{101})+(\ref{105})\\
+\qcdbggc
& &  {\rm from}\;(\ref{102})\\
+\qcdbgge
& &  {\rm from}\;(\ref{103})\\
+\frac{1}{4}\qcdbggac
& &  {\rm from}\;(\ref{106})+(\ref{110})\\
+\frac{1}{2}\qcdbggaa
& &  {\rm from}\;(\ref{104})+(\ref{106})\\
+\frac{1}{4}\qcdbggaf
& &  {\rm from}\;(\ref{107})+(\ref{111})\\
+\frac{1}{6}\qcdbggag
& &  {\rm from}\;(\ref{107})\\
+\left(\qcdbggu+\qcdbggv\right)
& &  {\rm from}\;(\ref{108})+(\ref{116})\\
+\left(\qcdbggx+\qcdbggy\right)
& &  {\rm from}\;(\ref{109})+(\ref{119})\\
+\left(\qcdbggz\right)
& &  {\rm from}\;(\ref{108})+(\ref{114})\\
+\left(\qcdbggw\right)
& &  {\rm from}\;(\ref{109})+(\ref{117})\\
+\frac{1}{2}\left(\qcdbggad\right)
& &  {\rm from}\;(\ref{112})\\
+\frac{1}{2}\left(\qcdbggae\right). & &  {\rm from}\;(\ref{113}) \eeas We indeed
confirm that the result is \be c_2^{\qcdzggb}=\sum_{|\Gamma|=2 \atop
{\bf res}(\Gamma)=\qcdzggb}\frac{\Gamma}{{\bf sym}(\Gamma)},\ee
the sum over all graphs at the given loop order, divided by their symmetry factors. This
confirms the Hochschild theorem.

Furthermore, we find that \bea \Delta^\prime(c^{\qcdzggb}_2) & = &
\left(2c_1^{\qcdzggc}+2c_1^{\qcdzggb}\right)\otimes \frac{1}{2}B_+^{\qcdaggb}(\One)\\
& + & \left(c_1^{\qcdzggd}+c_1^{\qcdzggb}\right)\otimes \frac{1}{2}B_+^{\qcdaggc}(\One)\\
& + & \left(2c_1^{\qcdzgff}+2c_1^{\qcdzff}\right)\otimes B_+^{\qcdagga}(\One)\\
& + & \left(2c_1^{\qcdzguu}+2c_1^{\qcdzuu}\right)\otimes
B_+^{\qcdaggd}(\One).\eea We now impose the Slavnov--Taylor
identity, which allows us to write the above as \be [2c^{\rm
coupl}_1-c_1^{\qcdzggb}]\otimes B_+^{1,\qcdzggb},\ee by expanding
(\ref{ST}) to order $g^2$. Vice versa, if we require that the
coproduct defines a sub Hopf algebra on the $c^r_j$, we reobtain the
Slavnov--Taylor identities \be
2c_1^{\qcdzggc}+2c_1^{\qcdzggb}=c_1^{\qcdzggd}+c_1^{\qcdzggb}=2c_1^{\qcdzgff}+2c_1^{\qcdzff}
=2c_1^{\qcdzguu}+2c_1^{\qcdzuu}.\ee Hence we recover the Slavnov
Taylor identities for the couplings from the above requirement.
Summarizing, we indeed find \be
\Delta^\prime(c_2^{\qcdzggb})=\left[2c_1^{\rm coupl}-c_1^{\qcdzggb}\right] \otimes
c_1^{\qcdzggb}.\ee

Note that the above indeed implies \be
bB_+^{1,\qcdzggb}\left(\Gamma^{\qcdzggb}[X_{\rm coupl}]^2\right)=0,\ee where
\be
B_+^{1,\qcdzggb}=\frac{1}{2}B_+^{\qcdaggb}+\frac{1}{2}B_+^{\qcdaggc}+B_+^{\qcdagga}+B_+^{\qcdaggd}.\ee
\subsection{Hochschild closedness}
Finally, it is instructive to see how the Hochschild closedness
comes about. Working out the coproduct on say the combination
$\frac{1}{6}\qcdbggag+\frac{1}{4}\qcdbggaf=:U$ we find \bea
\Delta(U) & = & U\otimes 1 +1\otimes U
+\frac{3}{6}\left(\qcdaggggj\otimes\qcdaggc\right)+
\frac{1}{4}\left(\qcdaggggj\otimes\qcdaggc\right)\nonumber\\
 & + & \frac{1}{4}\left(\qcdaggc\otimes \qcdaggc\right)\eea On the other
hand, looking at the definition of $c_1^{\qcdzggd}$, we find a mixed
term \be \frac{1}{2}\left( \qcdaggggj + \qcdaggggk + \qcdaggggl
\right) \otimes \frac{1}{2} \qcdaggc ,\ee and we now see why we
insist on a symmetric renormalization point.

Furthermore, we confirm \bea & &
\overbrace{n\left(\qcdaggc,\qcdaggggj,\qcdbggag\!\!\right)}^{3}=  \\
  & & \frac{\overbrace{{\bf
top}\left(\qcdaggc,\qcdaggggj,\qcdbggag\!\!\right)}^{2}\overbrace{{\bf
ram}\left(\qcdaggc,\qcdaggggj,\qcdbggag\!\!\right)}^{1}\overbrace{{\bf
sym}\left(\qcdbggag\!\!\right)}^{6}}{\underbrace{{\bf
sym}\left(\qcdaggc\right)}_{2}\underbrace{{\bf
sym}\left(\qcdaggggj\right)}_{2}},  \nonumber\eea as it must by
our definitions.
\end{fmffile}
\section{Discussion}
\begin{fmffile}{fmfG-tree}
\end{fmffile}
We have exhibited the inner workings of Hochschild cohomology in the
context of the Dyson--Schwinger equations of a generic non-abelian
gauge theories. As a first combinatorial exercise we related the
Slavnov--Taylor identities for the couplings to the very existence
of a sub Hopf algebra which is based on the sum of all graphs at a
given loop order. From \cite{Broadhurst:2000dq} we know that the
existence of this sub Hopf algebra is the first and crucial step
towards non-perturbative solutions of such equations. Further steps
in that direction are upcoming.

To prepare for this we finish the paper with a short discussion of
some further properties of our set-up. This is largely meant as an
outlook to upcoming results obtained by combining the Hopf algebra
approach to perturbation theory with the structure of
Dyson--Schwinger equations.
\subsection{Locality and Finiteness}
\begin{fmffile}{fmfG-feyn}
The first result concerns the proof of locality of counterterms and
finiteness of renormalizad Hopf algebra. The structure \be
\Gamma^{r}=1+\sum_k g^{2k}B_+^{k;r}(\Gamma^{r} [X_{\rm coupl}]^k)\ee
allows to prove locality of counterterms and finiteness of
renormalized Green function via induction over the augmentation
degree, involving nothing more than an elementary application of
Weinberg's theorem to primitive graphs \cite{Kreimer:2002rf}. It
unravels in that manner the source of equisingularity in the
corresponding Riemann--Hilbert correspondences
\cite{Connes:2004xb,CM}.
\end{fmffile}
For the DSE equations, this implies that we can define renormalized
Feynman rules via the choice of a suitable boundary condition. This
leads to an analytic study of the properties of the integral kernels
of $\phi(B_+^{k;r}(\One))$ to be given in future work. Furthermore,
the sub Hopf algebra of generators $c_k^{r}$ allows for recursions
similar to the ones employed in \cite{Broadhurst:2000dq}, relating
higher loop order amplitudes to products of lower loop order ones.
The most crucial ingredient of the non-perturbative methods employed
in that paper is now at our disposal for future work.

\subsection{Expansions in the conformal anomaly}
The form of the arguments $X_{r,k}=\Gamma^{r}X^k_{\rm coupl}$ allows
for a systematic expansion in the coefficients of the
$\beta$-function which relates the renormalization group to the
lower central series of the Lie algebra ${\cal L}$. Indeed, if the
$\beta$ function vanishes $X_{\rm coupl}$ is mapped under the
Feynman rules to a constant, and hence the resulting DSE become
linear, by inspection. One immediately confirms that the resulting
Hopf algebra structure is cocommutative, and the Lie algebra hence
abelian \cite{Bonn,Bergbauer:2005fb}. This should relate dilatations
in quantum field theory to the representation theory of that lower
central series. It will be interesting to compare the results here and more general in
\cite{review} with the ones in \cite{Korchemsky} from this viewpoint.
\subsection{Central Extensions}
The sub-Hopf algebras underlying the gauge theory theorem remain
invariant upon addition of new primitive elements - beyond the
one-loop level they obtain the form of a hierarchy of central
extensions, which clearly deserves further study. Indeed, if we were
to use only $B_+^{1,r}$ instead of the full series of Hochschild one
cocycles we would still obtain the same sub Hopf algebra. Thus, this
sub Hopf algebra and the structure of the DSEs is universal for a
chosen QFT in the sense of \cite{Bonn,Bergbauer:2005fb}.
\subsection{Radius of convergence}
The above structure ensures that the Green functions come as a
solution to a recursive equation which naturally provides one
primitive generator in each degree. This has remarkable consequences
for the radius of convergence when we express perturbation theory as
a series in the coefficients $c_k^r$, upon utilizing properties of
generating functions for recursive structures \cite{Karen}.
\subsection{Motivic picture} The primitives themselves relate
naturally to motivic theory \cite{BEK}. Each primitive generator is
transcendentally distinguished, with the one-loop iterated integral
providing the rational seed of the game. The relation to algebraic
geometry, motivic theory and mixed Hodge structures coming from QFT
as they slowly emerge in \cite{BEK,Bonn,CM} are an encouraging sign
of the deep mathematical underpinnings of local interacting quantum
fields.
\section*{Acknowledgments}
It is a pleasure to thank Christoph Bergbauer, David Broadhurst,
Kurusch Ebrahimi-Fard, Ivan Todorov and Karen Yeats for stimulating
discussions.


\end{document}

%% file: qcd.tex

\newcommand{\qcdzggb}{\raisebox{-0.0mm}{\begin{fmfchar}(6,3) \fmfkeep{qcd0gg2} \fmfpen{0.2mm}
\fmfset{curly_len}{0.7mm}
  \fmfleft{bin} 
  \fmf{gluon}{bin,bout}
  \fmfright{bout} 
\end{fmfchar}}}

\newcommand{\qcdzff}{\raisebox{-0.0mm}{\begin{fmfchar}(6,3) \fmfkeep{qcd0ff} \fmfpen{0.2mm}
\fmfset{curly_len}{0.7mm}\fmfset{arrow_len}{1.0mm}
\fmfset{dash_len}{0.8mm} \fmfset{dot_size}{1mm}
  \fmfleft{bin} 
  \fmf{fermion}{bin,bout}
  \fmfright{bout} 
\end{fmfchar}}}

\newcommand{\qcdzuu}{\raisebox{-0.0mm}{\begin{fmfchar}(6,3) \fmfkeep{qcd0uu} \fmfpen{0.2mm}
\fmfset{curly_len}{0.7mm}\fmfset{arrow_len}{1.0mm}
\fmfset{dash_len}{0.8mm} \fmfset{dot_size}{1mm}
  \fmfleft{bin} 
  \fmf{scalar}{bin,bout}
  \fmfright{bout} 
\end{fmfchar}}}

\newcommand{\qcdzggc}{\raisebox{-0.0mm}{\begin{fmfchar}(6,4) \fmfkeep{qcd0gg3} \fmfpen{0.2mm}
\fmfset{curly_len}{0.7mm}
  \fmfleft{bin} 
  \fmf{gluon}{bin,v}
  \fmf{gluon}{bout1,v,bout2}
  \fmfright{bout1,bout2} 
  \fmfv{decor.shape=circle,decor.filled=full,decor.size=0.7thick}{v}
\end{fmfchar}}}

\newcommand{\qcdzggd}{\raisebox{-0.0mm}{\begin{fmfchar}(6,4) \fmfkeep{qcd0gg4} \fmfpen{0.2mm}
\fmfset{curly_len}{0.7mm}
  \fmfleft{bin1,bin2} 
  \fmf{gluon}{bin1,v,bin2}
  \fmf{gluon}{bout1,v,bout2}
  \fmfright{bout1,bout2} 
  \fmfv{decor.shape=circle,decor.filled=full,decor.size=0.7thick}{v}
\end{fmfchar}}}

\newcommand{\qcdzgff}{\raisebox{-0.0mm}{\begin{fmfchar}(6,4) \fmfkeep{qcd0gff} \fmfpen{0.2mm}
\fmfset{curly_len}{0.7mm} \fmfset{arrow_len}{1.0mm}
\fmfset{dash_len}{0.8mm} \fmfset{dot_size}{1mm}

  \fmfleft{bin} 
  \fmf{gluon}{bin,v}
  \fmf{fermion}{bout1,v,bout2}
  \fmfright{bout1,bout2} 
  \fmfv{decor.shape=circle,decor.filled=full,decor.size=0.7thick}{v}
\end{fmfchar}}}

\newcommand{\qcdzguu}{\raisebox{-0.0mm}{\begin{fmfchar}(6,4) \fmfkeep{qcd0guu} \fmfpen{0.2mm}
\fmfset{curly_len}{0.7mm}  \fmfset{arrow_len}{1.0mm}
\fmfset{dash_len}{0.6mm} \fmfset{dot_size}{1mm}
  \fmfleft{bin} 
  \fmf{gluon}{bin,v}
  \fmf{scalar}{bout1,v,bout2}
  \fmfright{bout1,bout2} 
  \fmfv{decor.shape=circle,decor.filled=full,decor.size=0.7thick}{v}
\end{fmfchar}}}


\newcommand{\qcdaffa}{\raisebox{-2mm}{\begin{fmfchar}(9,6) \fmfkeep{qcd1ff1} \fmfpen{0.1mm}
\fmfset{curly_len}{0.4mm}\fmfset{arrow_len}{0.6mm}
\fmfset{dash_len}{0.4mm} \fmfset{dot_size}{1mm}
  \fmfleft{bin} 
  \fmf{fermion,tension=0.3}{v1,v2}
  \fmf{fermion}{bin,v1}
  \fmf{fermion}{v2,bout}
  \fmf{gluon,left,tension=0.3}{v1,v2}
  \fmfright{bout} 
  \fmfv{decor.shape=circle,decor.filled=full,decor.size=0.6thick}{v1,v2}
\end{fmfchar}}}


\newcommand{\qcdauua}{\raisebox{-2mm}{\begin{fmfchar}(9,6) \fmfkeep{qcd1uu1} \fmfpen{0.1mm}
\fmfset{curly_len}{0.4mm}\fmfset{arrow_len}{0.6mm}
\fmfset{dash_len}{0.4mm} \fmfset{dot_size}{1mm}
  \fmfleft{bin} 
  \fmf{scalar,tension=0.3}{v1,v2}
  \fmf{scalar}{bin,v1}
  \fmf{scalar}{v2,bout}
  \fmf{gluon,left,tension=0.3}{v1,v2}
  \fmfright{bout} 
  \fmfv{decor.shape=circle,decor.filled=full,decor.size=0.6thick}{v1,v2}
\end{fmfchar}}}


\newcommand{\qcdaggb}{\raisebox{-2mm}{\begin{fmfchar}(9,6) \fmfkeep{qcd1gg2} \fmfpen{0.1mm}
\fmfset{curly_len}{0.4mm}
  \fmfleft{bin} 
  \fmf{gluon}{bin,v1}
  \fmf{gluon}{v2,bout}
  \fmf{gluon,left,tension=.3}{v1,v2,v1}
  \fmfright{bout} 
  \fmfv{decor.shape=circle,decor.filled=full,decor.size=0.6thick}{v1,v2}
\end{fmfchar}}}

\newcommand{\qcdagga}{\raisebox{-2mm}{\begin{fmfchar}(9,6)
\fmfkeep{qcd1gg1} \fmfpen{0.1mm} \fmfset{curly_len}{0.4mm}
\fmfset{arrow_len}{2mm}
  \fmfleft{bin} 
  \fmf{gluon}{bin,v1}
  \fmf{gluon}{v2,bout}
  \fmf{fermion,left,tension=.3}{v1,v2,v1}
  \fmfright{bout} 
  \fmfv{decor.shape=circle,decor.filled=full,decor.size=0.6thick}{v1,v2}
\end{fmfchar}}}

\newcommand{\qcdaggd}{\raisebox{-2mm}{\begin{fmfchar}(9,6)
\fmfkeep{qcd1gg4} \fmfpen{0.1mm} \fmfset{curly_len}{0.4mm}
\fmfset{arrow_len}{2mm} \fmfset{dash_len}{1mm}
  \fmfleft{bin} 
  \fmf{gluon}{bin,v1}
  \fmf{gluon}{v2,bout}
  \fmf{scalar,left,tension=.3}{v1,v2,v1}
  \fmfright{bout} 
  \fmfv{decor.shape=circle,decor.filled=full,decor.size=0.6thick}{v1,v2}
\end{fmfchar}}}

\newcommand{\qcdaggc}{\raisebox{-2mm}{\begin{fmfchar}(9,6)
\fmfkeep{qcd1gg3} \fmfpen{0.1mm} \fmfset{curly_len}{0.4mm}
  \fmfleft{bin} 
  \fmf{gluon}{bin,v,v,bout}
  \fmfright{bout} 
  \fmfv{decor.shape=circle,decor.filled=full,decor.size=0.6thick}{v}
\end{fmfchar}}}

\newcommand{\qcdagffa}{\raisebox{-2mm}{\begin{fmfchar}(9,6)
\fmfkeep{qcd1gff1} \fmfpen{0.1mm} \fmfset{curly_len}{0.4mm}
\fmfset{arrow_len}{1mm} \fmfset{dash_len}{1mm}
  \fmfleft{bin} 
  \fmf{gluon}{bin,v}
  \fmf{gluon,tension=0.3}{v1,v2}
  \fmf{fermion,tension=0.3}{v2,v,v1}
  \fmf{fermion}{fout2,v2}
  \fmf{fermion}{v1,fout1}
  \fmfright{fout1,fout2} 
  \fmfv{decor.shape=circle,decor.filled=full,decor.size=0.6thick}{v,v1,v2}
\end{fmfchar}}}

\newcommand{\qcdagffb}{\raisebox{-2mm}{\begin{fmfchar}(9,6)
\fmfkeep{qcd1gff2} \fmfpen{0.1mm} \fmfset{curly_len}{0.4mm}
\fmfset{arrow_len}{1mm} \fmfset{dash_len}{1mm}
  \fmfleft{bin} 
  \fmf{gluon}{bin,v}
  \fmf{gluon,tension=0.3}{v1,v,v2}
  \fmf{fermion,tension=0.3}{v2,v1}
  \fmf{fermion}{fout2,v2}
  \fmf{fermion}{v1,fout1}
  \fmfright{fout1,fout2} 
  \fmfv{decor.shape=circle,decor.filled=full,decor.size=0.6thick}{v,v1,v2}
\end{fmfchar}}}

\newcommand{\qcdaguua}{\raisebox{-2mm}{\begin{fmfchar}(9,6)
\fmfkeep{qcd1guu1} \fmfpen{0.1mm} \fmfset{curly_len}{0.4mm}
\fmfset{arrow_len}{1mm} \fmfset{dash_len}{1mm}
  \fmfleft{bin} 
  \fmf{gluon}{bin,v}
  \fmf{gluon,tension=0.3}{v1,v2}
  \fmf{scalar,tension=0.3}{v2,v,v1}
  \fmf{scalar}{fout2,v2}
  \fmf{scalar}{v1,fout1}
  \fmfright{fout1,fout2} 
  \fmfv{decor.shape=circle,decor.filled=full,decor.size=0.6thick}{v,v1,v2}
\end{fmfchar}}}

\newcommand{\qcdaguub}{\raisebox{-2mm}{\begin{fmfchar}(9,6)
\fmfkeep{qcd1guu2} \fmfpen{0.1mm} \fmfset{curly_len}{0.4mm}
\fmfset{arrow_len}{1mm} \fmfset{dash_len}{1mm}
  \fmfleft{bin} 
  \fmf{gluon}{bin,v}
  \fmf{gluon,tension=0.3}{v1,v,v2}
  \fmf{scalar,tension=0.3}{v2,v1}
  \fmf{scalar}{fout2,v2}
  \fmf{scalar}{v1,fout1}
  \fmfright{fout1,fout2} 
  \fmfv{decor.shape=circle,decor.filled=full,decor.size=0.6thick}{v,v1,v2}
\end{fmfchar}}}

\newcommand{\qcdaggga}{\raisebox{-2mm}{\begin{fmfchar}(9,6)
\fmfkeep{qcd1ggg1} \fmfpen{0.1mm} \fmfset{curly_len}{0.4mm}
\fmfset{arrow_len}{1mm} \fmfset{dash_len}{1mm}
  \fmfleft{bin} 
  \fmf{gluon}{bin,v}
  \fmf{gluon}{v2,bout2}
  \fmf{gluon}{v1,bout1}
  \fmf{fermion,tension=0.3}{v2,v,v1,v2}
  \fmfright{bout1,bout2} 
  \fmfv{decor.shape=circle,decor.filled=full,decor.size=0.6thick}{v,v1,v2}
\end{fmfchar}}}

\newcommand{\qcdagggb}{\raisebox{-2mm}{\begin{fmfchar}(9,6)
\fmfkeep{qcd1ggg2} \fmfpen{0.1mm} \fmfset{curly_len}{0.4mm}
\fmfset{arrow_len}{1mm} \fmfset{dash_len}{1mm}
  \fmfleft{bin} 
  \fmf{gluon}{bin,v}
  \fmf{gluon}{v2,bout2}
  \fmf{gluon}{v1,bout1}
  \fmf{fermion,tension=0.3}{v2,v1,v,v2}
  \fmfright{bout1,bout2} 
  \fmfv{decor.shape=circle,decor.filled=full,decor.size=0.6thick}{v,v1,v2}
\end{fmfchar}}}

\newcommand{\qcdagggc}{\raisebox{-2mm}{\begin{fmfchar}(9,6)
\fmfkeep{qcd1ggg3} \fmfpen{0.1mm} \fmfset{curly_len}{0.4mm}
\fmfset{arrow_len}{1mm} \fmfset{dash_len}{1mm}
  \fmfleft{bin} 
  \fmf{gluon}{bin,v}
  \fmf{gluon}{v2,bout2}
  \fmf{gluon}{v1,bout1}
  \fmf{gluon,tension=0.3}{v2,v1,v,v2}
  \fmfright{bout1,bout2} 
  \fmfv{decor.shape=circle,decor.filled=full,decor.size=0.6thick}{v,v1,v2}
\end{fmfchar}}}

\newcommand{\qcdagggd}{\raisebox{-2mm}{\begin{fmfchar}(9,6)
\fmfkeep{qcd1ggg4} \fmfpen{0.1mm} \fmfset{curly_len}{0.4mm}
\fmfset{arrow_len}{1mm} \fmfset{dash_len}{1mm}
  \fmfleft{bin} 
  \fmf{gluon}{bin,v1}
  \fmf{gluon,left,tension=0.3}{v1,v2,v1}
  \fmf{gluon}{bout1,v2,bout2}
  \fmfright{bout1,bout2} 
  \fmfv{decor.shape=circle,decor.filled=full,decor.size=0.6thick}{v1,v2}
\end{fmfchar}}}

\newcommand{\qcdaggge}{\raisebox{-2mm}{\begin{fmfchar}(9,6)
\fmfkeep{qcd1ggg5} \fmfpen{0.1mm} \fmfset{curly_len}{0.4mm}
\fmfset{arrow_len}{1mm} \fmfset{dash_len}{1mm}
  \fmfleft{bin} 
  \fmf{gluon}{bin,v1}
  \fmf{gluon,left,tension=0.3}{v1,v2,v1}
  \fmf{gluon}{bout1,v1}
  \fmf{gluon}{bout2,v2}
  \fmfright{bout1,bout2} 
  \fmfv{decor.shape=circle,decor.filled=full,decor.size=0.6thick}{v1,v2}
\end{fmfchar}}}

\newcommand{\qcdagggf}{\raisebox{-2mm}{\begin{fmfchar}(9,6)
\fmfkeep{qcd1ggg6} \fmfpen{0.1mm} \fmfset{curly_len}{0.4mm}
\fmfset{arrow_len}{1mm} \fmfset{dash_len}{1mm}
  \fmfleft{bin} 
  \fmf{gluon}{bin,v1}
  \fmf{gluon,left,tension=0.3}{v1,v2,v1}
  \fmf{gluon}{bout1,v2}
  \fmf{gluon}{bout2,v1}
  \fmfright{bout1,bout2} 
  \fmfv{decor.shape=circle,decor.filled=full,decor.size=0.6thick}{v1,v2}
\end{fmfchar}}}

\newcommand{\qcdagggg}{\raisebox{-2mm}{\begin{fmfchar}(9,6)
\fmfkeep{qcd1ggg7} \fmfpen{0.1mm} \fmfset{curly_len}{0.4mm}
\fmfset{arrow_len}{1mm} \fmfset{dash_len}{1mm}
  \fmfleft{bin} 
  \fmf{gluon}{bin,v}
  \fmf{gluon}{v2,bout2}
  \fmf{gluon}{v1,bout1}
  \fmf{scalar,tension=0.3}{v2,v,v1,v2}
  \fmfright{bout1,bout2} 
  \fmfv{decor.shape=circle,decor.filled=full,decor.size=0.6thick}{v,v1,v2}
\end{fmfchar}}}

\newcommand{\qcdagggh}{\raisebox{-2mm}{\begin{fmfchar}(9,6)
\fmfkeep{qcd1ggg8} \fmfpen{0.1mm} \fmfset{curly_len}{0.4mm}
\fmfset{arrow_len}{1mm} \fmfset{dash_len}{1mm}
  \fmfleft{bin} 
  \fmf{gluon}{bin,v}
  \fmf{gluon}{v2,bout2}
  \fmf{gluon}{v1,bout1}
  \fmf{scalar,tension=0.3}{v2,v1,v,v2}
  \fmfright{bout1,bout2} 
  \fmfv{decor.shape=circle,decor.filled=full,decor.size=0.6thick}{v,v1,v2}
\end{fmfchar}}}


\newcommand{\qcdagggga}{\raisebox{-2mm}{\begin{fmfchar}(9,6)
\fmfkeep{qcd1gggg1} \fmfpen{0.1mm} \fmfset{curly_len}{0.4mm}
\fmfset{arrow_len}{1mm} \fmfset{dash_len}{1mm}
  \fmfleft{bin1,bin2} 
  \fmf{gluon}{bin1,v1}
  \fmf{gluon}{bin2,v2}
  \fmf{gluon}{v3,bout1}
  \fmf{gluon}{v4,bout2}
  \fmfright{bout1,bout2} 
  \fmf{gluon,tension=0.3}{v1,v2,v4,v3,v1}
  \fmfv{decor.shape=circle,decor.filled=full,decor.size=0.6thick}{v1,v2,v3,v4}
\end{fmfchar}}}

\newcommand{\qcdaggggb}{\raisebox{-2mm}{\begin{fmfchar}(9,6)
\fmfkeep{qcd1gggg2} \fmfpen{0.1mm} \fmfset{curly_len}{0.4mm}
\fmfset{arrow_len}{1mm} \fmfset{dash_len}{1mm}
  \fmfleft{bin1,bin2} 
  \fmf{gluon}{bin1,v1}
  \fmf{gluon}{bin2,v2}
  \fmf{gluon,rubout,tension=0}{v4,bout1}
  \fmf{gluon,rubout,tension=0}{v3,bout2}
  \fmf{phantom}{v3,bout1}
  \fmf{phantom}{v4,bout2}
  \fmfright{bout1,bout2} 
  \fmf{gluon,tension=0.3}{v1,v2,v4,v3,v1}
  \fmfv{decor.shape=circle,decor.filled=full,decor.size=0.6thick}{v1,v2,v3,v4}
\end{fmfchar}}}

\newcommand{\qcdaggggc}{\raisebox{-2mm}{\begin{fmfchar}(9,6)
\fmfkeep{qcd1gggg3} \fmfpen{0.1mm} \fmfset{curly_len}{0.4mm}
\fmfset{arrow_len}{1mm} \fmfset{dash_len}{1mm}
  \fmfleft{bin1,bin2} 
    \fmf{phantom}{bin1,v1}
      \fmf{phantom}{v3,bout1}
  \fmf{gluon,rubout,tension=0}{bin1,v3}
  \fmf{gluon}{bin2,v2}
  \fmf{gluon,rubout,tension=0}{v1,bout1}
  \fmf{gluon}{v4,bout2}
  \fmfright{bout1,bout2} 
  \fmf{gluon,tension=0.3}{v1,v2,v4,v3,v1}
  \fmfv{decor.shape=circle,decor.filled=full,decor.size=0.6thick}{v1,v2,v3,v4}
\end{fmfchar}}}

\newcommand{\qcdaggggd}{\raisebox{-2mm}{\begin{fmfchar}(9,6)
\fmfkeep{qcd1gggg4} \fmfpen{0.1mm} \fmfset{curly_len}{0.4mm}
\fmfset{arrow_len}{1mm} \fmfset{dash_len}{1mm}
  \fmfleft{bin1,bin2} 
  \fmf{gluon}{bin1,v1}
  \fmf{gluon}{bin2,v2}
  \fmf{gluon}{v,bout1}
  \fmf{gluon}{v,bout2}
  \fmfright{bout1,bout2} 
  \fmf{gluon,tension=0.3}{v1,v2,v,v1}
  \fmfv{decor.shape=circle,decor.filled=full,decor.size=0.6thick}{v1,v2,v}
\end{fmfchar}}}

\newcommand{\qcdagggge}{\raisebox{-2mm}{\begin{fmfchar}(9,6)
\fmfkeep{qcd1gggg5} \fmfpen{0.1mm} \fmfset{curly_len}{0.4mm}
\fmfset{arrow_len}{1mm} \fmfset{dash_len}{1mm}
  \fmfleft{bin1,bin2} 
  \fmf{gluon}{bin1,v}
  \fmf{gluon}{bin2,v1}
  \fmf{gluon}{v,bout1}
  \fmf{gluon}{v2,bout2}
  \fmfright{bout1,bout2} 
  \fmf{gluon,tension=0.3}{v1,v2,v,v1}
  \fmfv{decor.shape=circle,decor.filled=full,decor.size=0.6thick}{v1,v2,v}
\end{fmfchar}}}

\newcommand{\qcdaggggf}{\raisebox{-2mm}{\begin{fmfchar}(9,6)
\fmfkeep{qcd1gggg6} \fmfpen{0.1mm} \fmfset{curly_len}{0.4mm}
\fmfset{arrow_len}{1mm} \fmfset{dash_len}{1mm}
  \fmfleft{bin1,bin2} 
  \fmf{gluon}{bin1,v}
  \fmf{gluon}{bin2,v}
  \fmf{gluon}{v1,bout1}
  \fmf{gluon}{v2,bout2}
  \fmfright{bout1,bout2} 
  \fmf{gluon,tension=0.3}{v1,v2,v,v1}
  \fmfv{decor.shape=circle,decor.filled=full,decor.size=0.6thick}{v1,v2,v}
\end{fmfchar}}}

\newcommand{\qcdaggggg}{\raisebox{-2mm}{\begin{fmfchar}(9,6)
\fmfkeep{qcd1gggg7} \fmfpen{0.1mm} \fmfset{curly_len}{0.4mm}
\fmfset{arrow_len}{1mm} \fmfset{dash_len}{1mm}
  \fmfleft{bin1,bin2} 
  \fmf{gluon}{bin1,v1}
  \fmf{gluon}{bin2,v}
  \fmf{gluon}{v2,bout1}
  \fmf{gluon}{v,bout2}
  \fmfright{bout1,bout2} 
  \fmf{gluon,tension=0.3}{v1,v2,v,v1}
  \fmfv{decor.shape=circle,decor.filled=full,decor.size=0.6thick}{v1,v2,v}
\end{fmfchar}}}

\newcommand{\qcdaggggh}{\raisebox{-2mm}{\begin{fmfchar}(9,6)
\fmfkeep{qcd1gggg8} \fmfpen{0.1mm} \fmfset{curly_len}{0.4mm}
\fmfset{arrow_len}{1mm} \fmfset{dash_len}{1mm}
  \fmfleft{bin1,bin2} 
  \fmf{gluon}{bin1,v1}
  \fmf{gluon,rubout}{bin2,v}
  \fmf{gluon}{v,bout1}
  \fmf{gluon}{v2,bout2}
  \fmfright{bout1,bout2} 
  \fmf{gluon,left,tension=0.3}{v1,v2,v,v1}
  \fmfv{decor.shape=circle,decor.filled=full,decor.size=0.6thick}{v1,v2,v}
\end{fmfchar}}}

\newcommand{\qcdaggggi}{\raisebox{-2mm}{\begin{fmfchar}(9,6)
\fmfkeep{qcd1gggg9} \fmfpen{0.1mm} \fmfset{curly_len}{0.4mm}
\fmfset{arrow_len}{1mm} \fmfset{dash_len}{1mm}
  \fmfleft{bin1,bin2} 
  \fmf{gluon}{bin1,v}
  \fmf{gluon}{bin2,v1}
  \fmf{gluon}{v2,bout1}
  \fmf{gluon,rubout}{v,bout2}
  \fmfright{bout1,bout2} 
  \fmf{gluon,left,tension=0.3}{v1,v2,v,v1}
  \fmfv{decor.shape=circle,decor.filled=full,decor.size=0.6thick}{v1,v2,v}
\end{fmfchar}}}

\newcommand{\qcdaggggj}{\raisebox{-2mm}{\begin{fmfchar}(9,6)
\fmfkeep{qcd1gggg10} \fmfpen{0.1mm} \fmfset{curly_len}{0.4mm}
\fmfset{arrow_len}{1mm} \fmfset{dash_len}{1mm}
  \fmfleft{bin1,bin2} 
  \fmf{gluon}{bin1,v1}
  \fmf{gluon}{bin2,v1}
  \fmf{gluon}{v2,bout1}
  \fmf{gluon}{v2,bout2}
  \fmfright{bout1,bout2} 
  \fmf{gluon,left,tension=0.3}{v1,v2,v1}
  \fmfv{decor.shape=circle,decor.filled=full,decor.size=0.6thick}{v1,v2}
\end{fmfchar}}}

\newcommand{\qcdaggggk}{\raisebox{-2mm}{\begin{fmfchar}(9,6)
\fmfkeep{qcd1gggg11} \fmfpen{0.1mm} \fmfset{curly_len}{0.4mm}
\fmfset{arrow_len}{1mm} \fmfset{dash_len}{1mm}
  \fmfleft{bin1,bin2} 
  \fmf{gluon}{bin1,v1}
  \fmf{gluon}{bin2,v2}
  \fmf{gluon}{v1,bout1}
  \fmf{gluon}{v2,bout2}
  \fmfright{bout1,bout2} 
  \fmf{gluon,left,tension=0.3}{v1,v2,v1}
  \fmfv{decor.shape=circle,decor.filled=full,decor.size=0.6thick}{v1,v2}
\end{fmfchar}}}

\newcommand{\qcdaggggl}{\raisebox{-2mm}{\begin{fmfchar}(9,6)
\fmfkeep{qcd1gggg12} \fmfpen{0.1mm} \fmfset{curly_len}{0.4mm}
\fmfset{arrow_len}{1mm} \fmfset{dash_len}{1mm}
  \fmfleft{bin1,bin2} 
  \fmf{gluon}{bin1,v2}
  \fmf{gluon,tension=2}{bin2,v1}
  \fmf{gluon,left,rubout,tension=0.5}{v1,bout1}
  \fmf{gluon}{v2,bout2}
  \fmfright{bout1,bout2} 
  \fmf{gluon,left,tension=0.1}{v1,v2,v1}
  \fmfv{decor.shape=circle,decor.filled=full,decor.size=0.6thick}{v1,v2}
\end{fmfchar}}}

\newcommand{\qcdaggggm}{\raisebox{-2mm}{\begin{fmfchar}(9,6)
\fmfkeep{qcd1gggg13} \fmfpen{0.1mm} \fmfset{curly_len}{0.4mm}
\fmfset{arrow_len}{1mm} \fmfset{dash_len}{1mm}
  \fmfleft{bin1,bin2} 
  \fmf{gluon}{bin1,v1}
  \fmf{gluon}{bin2,v2}
  \fmf{gluon}{v3,bout1}
  \fmf{gluon}{v4,bout2}
  \fmfright{bout1,bout2} 
  \fmf{fermion,tension=0.3}{v1,v2,v4,v3,v1}
  \fmfv{decor.shape=circle,decor.filled=full,decor.size=0.6thick}{v1,v2,v3,v4}
\end{fmfchar}}}

\newcommand{\qcdaggggn}{\raisebox{-2mm}{\begin{fmfchar}(9,6)
\fmfkeep{qcd1gggg14} \fmfpen{0.1mm} \fmfset{curly_len}{0.4mm}
\fmfset{arrow_len}{1mm} \fmfset{dash_len}{1mm}
  \fmfleft{bin1,bin2} 
  \fmf{gluon}{bin1,v1}
  \fmf{gluon}{bin2,v2}
  \fmf{phantom}{v3,bout1}
  \fmf{phantom}{v4,bout2}
  \fmf{gluon,rubout,tension=0}{v4,bout1}
  \fmf{gluon,rubout,tension=0}{v3,bout2}
  \fmfright{bout1,bout2} 
  \fmf{fermion,tension=0.3}{v1,v2,v4,v3,v1}
  \fmfv{decor.shape=circle,decor.filled=full,decor.size=0.6thick}{v1,v2,v3,v4}
\end{fmfchar}}}

\newcommand{\qcdaggggo}{\raisebox{-2mm}{\begin{fmfchar}(9,6)
\fmfkeep{qcd1gggg15} \fmfpen{0.1mm} \fmfset{curly_len}{0.4mm}
\fmfset{arrow_len}{1mm} \fmfset{dash_len}{1mm}
  \fmfleft{bin1,bin2} 
  \fmf{gluon,rubout,tension=0}{bin1,v3}
  \fmf{gluon}{bin2,v2}
  \fmf{gluon,rubout,tension=0}{v1,bout1}
  \fmf{gluon}{v4,bout2}
  \fmf{phantom}{bin1,v1}
  \fmf{phantom}{v3,bout1}
  \fmfright{bout1,bout2} 
  \fmf{fermion,tension=0.3}{v1,v2,v4,v3,v1}
  \fmfv{decor.shape=circle,decor.filled=full,decor.size=0.6thick}{v1,v2,v3,v4}
\end{fmfchar}}}

\newcommand{\qcdaggggp}{\raisebox{-2mm}{\begin{fmfchar}(9,6)
\fmfkeep{qcd1gggg16} \fmfpen{0.1mm} \fmfset{curly_len}{0.4mm}
\fmfset{arrow_len}{1mm} \fmfset{dash_len}{1mm}
  \fmfleft{bin1,bin2} 
  \fmf{gluon}{bin1,v1}
  \fmf{gluon}{bin2,v2}
  \fmf{gluon}{v3,bout1}
  \fmf{gluon}{v4,bout2}
  \fmfright{bout1,bout2} 
  \fmf{fermion,tension=0.3}{v1,v3,v4,v2,v1}
  \fmfv{decor.shape=circle,decor.filled=full,decor.size=0.6thick}{v1,v2,v3,v4}
\end{fmfchar}}}

\newcommand{\qcdaggggq}{\raisebox{-2mm}{\begin{fmfchar}(9,6)
\fmfkeep{qcd1gggg17} \fmfpen{0.1mm} \fmfset{curly_len}{0.4mm}
\fmfset{arrow_len}{1mm} \fmfset{dash_len}{1mm}
  \fmfleft{bin1,bin2} 
  \fmf{gluon}{bin1,v1}
  \fmf{gluon}{bin2,v2}
  \fmf{gluon,rubout,tension=0}{v4,bout1}
  \fmf{gluon,rubout,tension=0}{v3,bout2}
  \fmf{phantom}{v3,bout1}
  \fmf{phantom}{v4,bout2}
  \fmfright{bout1,bout2} 
  \fmf{fermion,tension=0.3}{v1,v3,v4,v2,v1}
  \fmfv{decor.shape=circle,decor.filled=full,decor.size=0.6thick}{v1,v2,v3,v4}
\end{fmfchar}}}

\newcommand{\qcdaggggr}{\raisebox{-2mm}{\begin{fmfchar}(9,6)
\fmfkeep{qcd1gggg18} \fmfpen{0.1mm} \fmfset{curly_len}{0.4mm}
\fmfset{arrow_len}{1mm} \fmfset{dash_len}{1mm}
  \fmfleft{bin1,bin2} 
  \fmfleft{bin1,bin2} 
  \fmf{gluon,rubout,tension=0}{bin1,v3}
  \fmf{gluon}{bin2,v2}
  \fmf{gluon,rubout,tension=0}{v1,bout1}
  \fmf{gluon}{v4,bout2}
  \fmf{phantom}{bin1,v1}
  \fmf{phantom}{v3,bout1}
  \fmfright{bout1,bout2} 
  \fmf{fermion,tension=0.3}{v1,v3,v4,v2,v1}
  \fmfv{decor.shape=circle,decor.filled=full,decor.size=0.6thick}{v1,v2,v3,v4}
\end{fmfchar}}}

\newcommand{\qcdaggggs}{\raisebox{-2mm}{\begin{fmfchar}(9,6)
\fmfkeep{qcd1gggg19} \fmfpen{0.1mm} \fmfset{curly_len}{0.4mm}
\fmfset{arrow_len}{1mm} \fmfset{dash_len}{1mm}
  \fmfleft{bin1,bin2} 
  \fmf{gluon}{bin1,v1}
  \fmf{gluon}{bin2,v2}
  \fmf{gluon}{v3,bout1}
  \fmf{gluon}{v4,bout2}
  \fmfright{bout1,bout2} 
  \fmf{scalar,tension=0.3}{v1,v2,v4,v3,v1}
  \fmfv{decor.shape=circle,decor.filled=full,decor.size=0.6thick}{v1,v2,v3,v4}
\end{fmfchar}}}

\newcommand{\qcdaggggt}{\raisebox{-2mm}{\begin{fmfchar}(9,6)
\fmfkeep{qcd1gggg20} \fmfpen{0.1mm} \fmfset{curly_len}{0.4mm}
\fmfset{arrow_len}{1mm} \fmfset{dash_len}{1mm}
  \fmfleft{bin1,bin2} 
  \fmf{gluon}{bin1,v1}
  \fmf{gluon}{bin2,v2}
  \fmf{gluon,rubout,tension=0}{v4,bout1}
  \fmf{gluon,rubout,tension=0}{v3,bout2}
  \fmf{phantom}{v3,bout1}
  \fmf{phantom}{v4,bout2}
  \fmfright{bout1,bout2} 
  \fmf{scalar,tension=0.3}{v1,v2,v4,v3,v1}
  \fmfv{decor.shape=circle,decor.filled=full,decor.size=0.6thick}{v1,v2,v3,v4}
\end{fmfchar}}}

\newcommand{\qcdaggggu}{\raisebox{-2mm}{\begin{fmfchar}(9,6)
\fmfkeep{qcd1gggg21} \fmfpen{0.1mm} \fmfset{curly_len}{0.4mm}
\fmfset{arrow_len}{1mm} \fmfset{dash_len}{1mm}
  \fmfleft{bin1,bin2} 
  \fmf{gluon,rubout,tension=0}{bin1,v3}
  \fmf{gluon}{bin2,v2}
  \fmf{gluon,rubout,tension=0}{v1,bout1}
  \fmf{gluon}{v4,bout2}
  \fmf{phantom}{bin1,v1}
  \fmf{phantom}{v3,bout1}
  \fmfright{bout1,bout2} 
  \fmf{scalar,tension=0.3}{v1,v2,v4,v3,v1}
  \fmfv{decor.shape=circle,decor.filled=full,decor.size=0.6thick}{v1,v2,v3,v4}
\end{fmfchar}}}

\newcommand{\qcdaggggv}{\raisebox{-2mm}{\begin{fmfchar}(9,6)
\fmfkeep{qcd1gggg22} \fmfpen{0.1mm} \fmfset{curly_len}{0.4mm}
\fmfset{arrow_len}{1mm} \fmfset{dash_len}{1mm}
  \fmfleft{bin1,bin2} 
  \fmf{gluon}{bin1,v1}
  \fmf{gluon}{bin2,v2}
  \fmf{gluon}{v3,bout1}
  \fmf{gluon}{v4,bout2}
  \fmfright{bout1,bout2} 
  \fmf{scalar,tension=0.3}{v1,v3,v4,v2,v1}
  \fmfv{decor.shape=circle,decor.filled=full,decor.size=0.6thick}{v1,v2,v3,v4}
\end{fmfchar}}}

\newcommand{\qcdaggggw}{\raisebox{-2mm}{\begin{fmfchar}(9,6)
\fmfkeep{qcd1gggg23} \fmfpen{0.1mm} \fmfset{curly_len}{0.4mm}
\fmfset{arrow_len}{1mm} \fmfset{dash_len}{1mm}
  \fmfleft{bin1,bin2} 
  \fmf{gluon}{bin1,v1}
  \fmf{gluon}{bin2,v2}
  \fmf{gluon,rubout,tension=0}{v4,bout1}
  \fmf{gluon,rubout,tension=0}{v3,bout2}
  \fmf{phantom}{v3,bout1}
  \fmf{phantom}{v4,bout2}
  \fmfright{bout1,bout2} 
  \fmf{scalar,tension=0.3}{v1,v3,v4,v2,v1}
  \fmfv{decor.shape=circle,decor.filled=full,decor.size=0.6thick}{v1,v2,v3,v4}
\end{fmfchar}}}

\newcommand{\qcdaggggx}{\raisebox{-2mm}{\begin{fmfchar}(9,6)
\fmfkeep{qcd1gggg24} \fmfpen{0.1mm} \fmfset{curly_len}{0.4mm}
\fmfset{arrow_len}{1mm} \fmfset{dash_len}{1mm}
  \fmfleft{bin1,bin2} 
  \fmfleft{bin1,bin2} 
  \fmf{gluon,rubout,tension=0}{bin1,v3}
  \fmf{gluon}{bin2,v2}
  \fmf{gluon,rubout,tension=0}{v1,bout1}
  \fmf{gluon}{v4,bout2}
  \fmf{phantom}{bin1,v1}
  \fmf{phantom}{v3,bout1}
  \fmfright{bout1,bout2} 
  \fmf{scalar,tension=0.3}{v1,v3,v4,v2,v1}
  \fmfv{decor.shape=circle,decor.filled=full,decor.size=0.6thick}{v1,v2,v3,v4}
\end{fmfchar}}}


\newcommand{\qcdbgga}{\raisebox{-5mm}{\begin{fmfchar}(12,9) \fmfkeep{qcd2gg1} \fmfpen{0.1mm}
\fmfset{curly_len}{0.4mm} \fmfset{arrow_len}{1mm}
\fmfset{dash_len}{1mm} \fmfset{dot_size}{1mm}
   \fmfleft{bin} 
   \fmfright{bout} 
   \fmf{gluon}{bin,v1}
   \fmf{gluon}{v2,bout}
   \fmf{gluon,left,tension=0.1}{v1,v2}
   \fmf{gluon,tension=0.1}{v2,v3}
   \fmf{gluon,tension=0.1}{v4,v1}
   \fmf{gluon,left,tension=0.05}{v4,v3,v4}
  \fmfv{decor.shape=circle,decor.filled=full,decor.size=0.6thick}{v1,v2,v3,v4}
\end{fmfchar}}}

\newcommand{\qcdbggb}{\raisebox{-1mm}{\begin{fmfchar}(12,9) \fmfkeep{qcd2gg2} \fmfpen{0.1mm}
\fmfset{curly_len}{0.4mm} \fmfset{arrow_len}{1mm}
\fmfset{dash_len}{1mm} \fmfset{dot_size}{1mm}
   \fmfleft{bin} 
   \fmfright{bout} 
   \fmf{gluon}{bin,v1}
   \fmf{gluon}{v2,bout}
   \fmf{gluon,left,tension=0.1}{v2,v1}
   \fmf{gluon,tension=0.1}{v2,v3}
   \fmf{gluon,tension=0.1}{v4,v1}
   \fmf{gluon,left,tension=0.05}{v4,v3,v4}
  \fmfv{decor.shape=circle,decor.filled=full,decor.size=0.6thick}{v1,v2,v3,v4}
\end{fmfchar}}}

\newcommand{\qcdbggc}{\raisebox{-1mm}{\begin{fmfchar}(12,9) \fmfkeep{qcd2gg3} \fmfpen{0.1mm}
\fmfset{curly_len}{0.4mm} \fmfset{arrow_len}{1mm}
\fmfset{dash_len}{1mm} \fmfset{dot_size}{1mm}
   \fmfleft{bin} 
   \fmfright{bout} 
   \fmf{gluon}{bin,v1}
   \fmf{gluon}{v2,bout}
   \fmf{gluon,left,tension=0.1}{v2,v1}
   \fmf{gluon,tension=0.1}{v2,v3}
   \fmf{gluon,tension=0.1}{v4,v1}
   \fmf{fermion,left,tension=0.05}{v4,v3,v4}
  \fmfv{decor.shape=circle,decor.filled=full,decor.size=0.6thick}{v1,v2,v3,v4}
\end{fmfchar}}}

\newcommand{\qcdbggd}{\raisebox{-5mm}{\begin{fmfchar}(12,9) \fmfkeep{qcd2gg4} \fmfpen{0.1mm}
\fmfset{curly_len}{0.4mm} \fmfset{arrow_len}{1mm}
\fmfset{dash_len}{1mm} \fmfset{dot_size}{1mm}
   \fmfleft{bin} 
   \fmfright{bout} 
   \fmf{gluon}{bin,v1}
   \fmf{gluon}{v2,bout}
   \fmf{gluon,left,tension=0.1}{v1,v2}
   \fmf{gluon,tension=0.1}{v2,v3}
   \fmf{gluon,tension=0.1}{v4,v1}
   \fmf{fermion,left,tension=0.05}{v4,v3,v4}
  \fmfv{decor.shape=circle,decor.filled=full,decor.size=0.6thick}{v1,v2,v3,v4}
\end{fmfchar}}}

\newcommand{\qcdbgge}{\raisebox{-1mm}{\begin{fmfchar}(12,9) \fmfkeep{qcd2gg5} \fmfpen{0.1mm}
\fmfset{curly_len}{0.4mm} \fmfset{arrow_len}{1mm}
\fmfset{dash_len}{1mm} \fmfset{dot_size}{1mm}
   \fmfleft{bin} 
   \fmfright{bout} 
   \fmf{gluon}{bin,v1}
   \fmf{gluon}{v2,bout}
   \fmf{gluon,left,tension=0.1}{v2,v1}
   \fmf{gluon,tension=0.1}{v2,v3}
   \fmf{gluon,tension=0.1}{v4,v1}
   \fmf{scalar,left,tension=0.05}{v4,v3,v4}
  \fmfv{decor.shape=circle,decor.filled=full,decor.size=0.6thick}{v1,v2,v3,v4}
\end{fmfchar}}}

\newcommand{\qcdbggf}{\raisebox{-5mm}{\begin{fmfchar}(12,9) \fmfkeep{qcd2gg6} \fmfpen{0.1mm}
\fmfset{curly_len}{0.4mm} \fmfset{arrow_len}{1mm}
\fmfset{dash_len}{1mm} \fmfset{dot_size}{1mm}
   \fmfleft{bin} 
   \fmfright{bout} 
   \fmf{gluon}{bin,v1}
   \fmf{gluon}{v2,bout}
   \fmf{gluon,left,tension=0.1}{v1,v2}
   \fmf{gluon,tension=0.1}{v2,v3}
   \fmf{gluon,tension=0.1}{v4,v1}
   \fmf{scalar,left,tension=0.05}{v4,v3,v4}
  \fmfv{decor.shape=circle,decor.filled=full,decor.size=0.6thick}{v1,v2,v3,v4}
\end{fmfchar}}}

\newcommand{\qcdbggv}{\raisebox{-5mm}{\begin{fmfchar}(12,9) \fmfkeep{qcd2gg22} \fmfpen{0.1mm}
\fmfset{curly_len}{0.4mm} \fmfset{arrow_len}{1mm}
\fmfset{dash_len}{1mm} \fmfset{dot_size}{1mm}
  \fmfleft{bin} 
  \fmfright{bout} 
     \fmf{gluon}{bin,v1}
   \fmf{gluon}{v2,bout}
   \fmf{fermion,left,tension=0.1}{v1,v2}
   \fmf{fermion,tension=0.1}{v2,v3}
   \fmf{fermion,tension=0.1}{v4,v1}
   \fmf{fermion,left,tension=0.05}{v3,v4}
   \fmf{gluon,left,tension=0.05}{v4,v3}
  \fmfv{decor.shape=circle,decor.filled=full,decor.size=0.6thick}{v1,v2,v3,v4}
\end{fmfchar}}}

\newcommand{\qcdbggu}{\raisebox{-1mm}{\begin{fmfchar}(12,9) \fmfkeep{qcd2gg21} \fmfpen{0.1mm}
\fmfset{curly_len}{0.4mm} \fmfset{arrow_len}{1mm}
\fmfset{dash_len}{1mm} \fmfset{dot_size}{1mm}
   \fmfleft{bin} 
   \fmfright{bout} 
        \fmf{gluon}{bin,v1}
   \fmf{gluon}{v2,bout}
      \fmf{fermion,left,tension=0.1}{v2,v1}
   \fmf{fermion,tension=0.1}{v3,v2}
   \fmf{fermion,tension=0.1}{v1,v4}
   \fmf{fermion,right,tension=0.05}{v4,v3}
   \fmf{gluon,left,tension=0.05}{v4,v3}
  \fmfv{decor.shape=circle,decor.filled=full,decor.size=0.6thick}{v1,v2,v3,v4}
\end{fmfchar}}}

\newcommand{\qcdbggy}{\raisebox{-5mm}{\begin{fmfchar}(12,9) \fmfkeep{qcd2gg25} \fmfpen{0.1mm}
\fmfset{curly_len}{0.4mm} \fmfset{arrow_len}{1mm}
\fmfset{dash_len}{1mm} \fmfset{dot_size}{1mm}
  \fmfleft{bin} 
  \fmfright{bout} 
     \fmf{gluon}{bin,v1}
   \fmf{gluon}{v2,bout}
   \fmf{scalar,left,tension=0.1}{v1,v2}
   \fmf{scalar,tension=0.1}{v2,v3}
   \fmf{scalar,tension=0.1}{v4,v1}
   \fmf{scalar,left,tension=0.05}{v3,v4}
   \fmf{gluon,left,tension=0.05}{v4,v3}
  \fmfv{decor.shape=circle,decor.filled=full,decor.size=0.6thick}{v1,v2,v3,v4}
\end{fmfchar}}}

\newcommand{\qcdbggx}{\raisebox{-1mm}{\begin{fmfchar}(12,9) \fmfkeep{qcd2gg24} \fmfpen{0.1mm}
\fmfset{curly_len}{0.4mm} \fmfset{arrow_len}{1mm}
\fmfset{dash_len}{1mm} \fmfset{dot_size}{1mm}
   \fmfleft{bin} 
   \fmfright{bout} 
        \fmf{gluon}{bin,v1}
   \fmf{gluon}{v2,bout}
      \fmf{scalar,left,tension=0.1}{v2,v1}
   \fmf{scalar,tension=0.1}{v3,v2}
   \fmf{scalar,tension=0.1}{v1,v4}
   \fmf{scalar,right,tension=0.05}{v4,v3}
   \fmf{gluon,left,tension=0.05}{v4,v3}
  \fmfv{decor.shape=circle,decor.filled=full,decor.size=0.6thick}{v1,v2,v3,v4}
\end{fmfchar}}}

\newcommand{\qcdbggg}{\raisebox{-3mm}{\begin{fmfchar}(12,9) \fmfkeep{qcd2gg7} \fmfpen{0.1mm}
\fmfset{curly_len}{0.4mm} \fmfset{arrow_len}{1mm}
\fmfset{dash_len}{1mm} \fmfset{dot_size}{1mm}
  \fmfsurroundn{v}{4}
  \fmf{phantom}{v1,w1}
  \fmf{phantom,tension=2}{v2,w2}
  \fmf{phantom}{v3,w3}
  \fmf{phantom,tension=2}{v4,w4}
  \fmf{gluon}{w1,w2}
  \fmf{gluon}{w2,w3}
  \fmf{gluon}{w3,w4}
  \fmf{gluon}{w4,w1}
  \fmf{gluon,tension=0.3}{w2,w4}
  \fmf{gluon}{w1,v1}
  \fmf{gluon}{v3,w3}
  \fmfv{decor.shape=circle,decor.filled=full,decor.size=0.6thick}{w1,w2,w3,w4}
\end{fmfchar}}}

\newcommand{\qcdbggw}{\raisebox{-3mm}{\begin{fmfchar}(12,9) \fmfkeep{qcd2gg23} \fmfpen{0.1mm}
\fmfset{curly_len}{0.4mm} \fmfset{arrow_len}{1mm}
\fmfset{dash_len}{1mm} \fmfset{dot_size}{1mm}
  \fmfsurroundn{v}{4}
  \fmf{phantom}{v1,w1}
  \fmf{phantom,tension=2}{v2,w2}
  \fmf{phantom}{v3,w3}
  \fmf{phantom,tension=2}{v4,w4}
  \fmf{scalar}{w1,w2}
  \fmf{scalar}{w2,w3}
  \fmf{scalar}{w3,w4}
  \fmf{scalar}{w4,w1}
  \fmf{gluon,tension=0.3}{w2,w4}
  \fmf{gluon}{w1,v1}
  \fmf{gluon}{v3,w3}
  \fmfv{decor.shape=circle,decor.filled=full,decor.size=0.6thick}{w1,w2,w3,w4}
\end{fmfchar}}}

\newcommand{\qcdbggz}{\raisebox{-3mm}{\begin{fmfchar}(12,9) \fmfkeep{qcd2gg26} \fmfpen{0.1mm}
\fmfset{curly_len}{0.4mm} \fmfset{arrow_len}{1mm}
\fmfset{dash_len}{1mm} \fmfset{dot_size}{1mm}
  \fmfsurroundn{v}{4}
  \fmf{phantom}{v1,w1}
  \fmf{phantom,tension=2}{v2,w2}
  \fmf{phantom}{v3,w3}
  \fmf{phantom,tension=2}{v4,w4}
  \fmf{fermion}{w1,w2}
  \fmf{fermion}{w2,w3}
  \fmf{fermion}{w3,w4}
  \fmf{fermion}{w4,w1}
  \fmf{gluon,tension=0.3}{w2,w4}
  \fmf{gluon}{w1,v1}
  \fmf{gluon}{v3,w3}
  \fmfv{decor.shape=circle,decor.filled=full,decor.size=0.6thick}{w1,w2,w3,w4}
\end{fmfchar}}}

\newcommand{\qcdbggj}{\raisebox{-3mm}{\begin{fmfchar}(12,9) \fmfkeep{qcd2gg10} \fmfpen{0.1mm}
\fmfset{curly_len}{0.4mm} \fmfset{arrow_len}{1mm}
\fmfset{dash_len}{1mm} \fmfset{dot_size}{1mm}
  \fmfsurroundn{v}{4}
  \fmf{phantom}{v1,w1}
  \fmf{phantom,tension=2}{v2,w2}
  \fmf{phantom}{v3,w3}
  \fmf{phantom,tension=2}{v4,w4}
  \fmf{fermion}{w1,w2}
  \fmf{gluon}{w2,w3}
  \fmf{gluon}{w3,w4}
  \fmf{fermion}{w4,w1}
  \fmf{fermion,tension=0.3}{w2,w4}
  \fmf{gluon}{w1,v1}
  \fmf{gluon}{v3,w3}
  \fmfv{decor.shape=circle,decor.filled=full,decor.size=0.6thick}{w1,w2,w3,w4}
\end{fmfchar}}}

\newcommand{\qcdbggk}{\raisebox{-3mm}{\begin{fmfchar}(12,9) \fmfkeep{qcd2gg11} \fmfpen{0.1mm}
\fmfset{curly_len}{0.4mm} \fmfset{arrow_len}{1mm}
\fmfset{dash_len}{1mm} \fmfset{dot_size}{1mm}
  \fmfsurroundn{v}{4}
  \fmf{phantom}{v1,w1}
  \fmf{phantom,tension=2}{v2,w2}
  \fmf{phantom}{v3,w3}
  \fmf{phantom,tension=2}{v4,w4}
  \fmf{fermion}{w2,w1}
  \fmf{gluon}{w2,w3}
  \fmf{gluon}{w3,w4}
  \fmf{fermion}{w1,w4}
  \fmf{fermion,tension=0.3}{w4,w2}
  \fmf{gluon}{w1,v1}
  \fmf{gluon}{v3,w3}
  \fmfv{decor.shape=circle,decor.filled=full,decor.size=0.6thick}{w1,w2,w3,w4}
\end{fmfchar}}}

\newcommand{\qcdbggh}{\raisebox{-3mm}{\begin{fmfchar}(12,9) \fmfkeep{qcd2gg8} \fmfpen{0.1mm}
\fmfset{curly_len}{0.4mm} \fmfset{arrow_len}{1mm}
\fmfset{dash_len}{1mm} \fmfset{dot_size}{1mm}
  \fmfsurroundn{v}{4}
  \fmf{phantom}{v1,w1}
  \fmf{phantom,tension=2}{v2,w2}
  \fmf{phantom}{v3,w3}
  \fmf{phantom,tension=2}{v4,w4}
  \fmf{gluon}{w1,w2}
  \fmf{fermion}{w2,w3}
  \fmf{fermion}{w3,w4}
  \fmf{gluon}{w4,w1}
  \fmf{fermion,tension=0.3}{w4,w2}
  \fmf{gluon}{w1,v1}
  \fmf{gluon}{v3,w3}
  \fmfv{decor.shape=circle,decor.filled=full,decor.size=0.6thick}{w1,w2,w3,w4}
\end{fmfchar}}}

\newcommand{\qcdbggi}{\raisebox{-3mm}{\begin{fmfchar}(12,9) \fmfkeep{qcd2gg9} \fmfpen{0.1mm}
\fmfset{curly_len}{0.4mm} \fmfset{arrow_len}{1mm}
\fmfset{dash_len}{1mm} \fmfset{dot_size}{1mm}
  \fmfsurroundn{v}{4}
  \fmf{phantom}{v1,w1}
  \fmf{phantom,tension=2}{v2,w2}
  \fmf{phantom}{v3,w3}
  \fmf{phantom,tension=2}{v4,w4}
  \fmf{gluon}{w1,w2}
  \fmf{fermion}{w3,w2}
  \fmf{fermion}{w4,w3}
  \fmf{gluon}{w4,w1}
  \fmf{fermion,tension=0.3}{w2,w4}
  \fmf{gluon}{w1,v1}
  \fmf{gluon}{v3,w3}
  \fmfv{decor.shape=circle,decor.filled=full,decor.size=0.6thick}{w1,w2,w3,w4}
\end{fmfchar}}}

\newcommand{\qcdbggm}{\raisebox{-3mm}{\begin{fmfchar}(12,9) \fmfkeep{qcd2gg13} \fmfpen{0.1mm}
\fmfset{curly_len}{0.4mm} \fmfset{arrow_len}{1mm}
\fmfset{dash_len}{1mm} \fmfset{dot_size}{1mm}
  \fmfsurroundn{v}{4}
  \fmf{phantom}{v1,w1}
  \fmf{phantom,tension=2}{v2,w2}
  \fmf{phantom}{v3,w3}
  \fmf{phantom,tension=2}{v4,w4}
  \fmf{scalar}{w1,w2}
  \fmf{gluon}{w2,w3}
  \fmf{gluon}{w3,w4}
  \fmf{scalar}{w4,w1}
  \fmf{scalar,tension=0.3}{w2,w4}
  \fmf{gluon}{w1,v1}
  \fmf{gluon}{v3,w3}
  \fmfv{decor.shape=circle,decor.filled=full,decor.size=0.6thick}{w1,w2,w3,w4}
\end{fmfchar}}}

\newcommand{\qcdbggo}{\raisebox{-3mm}{\begin{fmfchar}(12,9) \fmfkeep{qcd2gg15} \fmfpen{0.1mm}
\fmfset{curly_len}{0.4mm} \fmfset{arrow_len}{1mm}
\fmfset{dash_len}{1mm} \fmfset{dot_size}{1mm}
  \fmfsurroundn{v}{4}
  \fmf{phantom}{v1,w1}
  \fmf{phantom,tension=2}{v2,w2}
  \fmf{phantom}{v3,w3}
  \fmf{phantom,tension=2}{v4,w4}
  \fmf{scalar}{w2,w1}
  \fmf{gluon}{w2,w3}
  \fmf{gluon}{w3,w4}
  \fmf{scalar}{w1,w4}
  \fmf{scalar,tension=0.3}{w4,w2}
  \fmf{gluon}{w1,v1}
  \fmf{gluon}{v3,w3}
  \fmfv{decor.shape=circle,decor.filled=full,decor.size=0.6thick}{w1,w2,w3,w4}
\end{fmfchar}}}

\newcommand{\qcdbggl}{\raisebox{-3mm}{\begin{fmfchar}(12,9) \fmfkeep{qcd2gg12} \fmfpen{0.1mm}
\fmfset{curly_len}{0.4mm} \fmfset{arrow_len}{1mm}
\fmfset{dash_len}{1mm} \fmfset{dot_size}{1mm}
  \fmfsurroundn{v}{4}
  \fmf{phantom}{v1,w1}
  \fmf{phantom,tension=2}{v2,w2}
  \fmf{phantom}{v3,w3}
  \fmf{phantom,tension=2}{v4,w4}
  \fmf{gluon}{w1,w2}
  \fmf{scalar}{w2,w3}
  \fmf{scalar}{w3,w4}
  \fmf{gluon}{w4,w1}
  \fmf{scalar,tension=0.3}{w4,w2}
  \fmf{gluon}{w1,v1}
  \fmf{gluon}{v3,w3}
  \fmfv{decor.shape=circle,decor.filled=full,decor.size=0.6thick}{w1,w2,w3,w4}
\end{fmfchar}}}

\newcommand{\qcdbggn}{\raisebox{-3mm}{\begin{fmfchar}(12,9) \fmfkeep{qcd2gg14} \fmfpen{0.1mm}
\fmfset{curly_len}{0.4mm} \fmfset{arrow_len}{1mm}
\fmfset{dash_len}{1mm} \fmfset{dot_size}{1mm}
  \fmfsurroundn{v}{4}
  \fmf{phantom}{v1,w1}
  \fmf{phantom,tension=2}{v2,w2}
  \fmf{phantom}{v3,w3}
  \fmf{phantom,tension=2}{v4,w4}
  \fmf{gluon}{w1,w2}
  \fmf{scalar}{w3,w2}
  \fmf{scalar}{w4,w3}
  \fmf{gluon}{w4,w1}
  \fmf{scalar,tension=0.3}{w2,w4}
  \fmf{gluon}{w1,v1}
  \fmf{gluon}{v3,w3}
  \fmfv{decor.shape=circle,decor.filled=full,decor.size=0.6thick}{w1,w2,w3,w4}
\end{fmfchar}}}

\newcommand{\qcdbggq}{\raisebox{-4mm}{\begin{fmfchar}(12,9) \fmfkeep{qcd2gg17} \fmfpen{0.1mm}
\fmfset{curly_len}{0.4mm} \fmfset{arrow_len}{1mm}
\fmfset{dash_len}{1mm} \fmfset{dot_size}{1mm}
  \fmfsurroundn{v}{4}
  \fmf{phantom}{v1,w1}
  \fmf{phantom,tension=2}{v2,w2}
  \fmf{phantom}{v3,w3}
  \fmf{gluon,left}{w2,w1}
  \fmf{gluon,left,tension=0.5}{w2,w3}
  \fmf{gluon,left}{w1,w3}
  \fmf{gluon,left,tension=0.5}{w3,w2}
  \fmf{gluon}{w1,v1}
  \fmf{gluon}{v3,w3}
  \fmfv{decor.shape=circle,decor.filled=full,decor.size=0.6thick}{w1,w2,w3}
\end{fmfchar}}}

\newcommand{\qcdbggr}{\raisebox{-2mm}{\begin{fmfchar}(12,9) \fmfkeep{qcd2gg18} \fmfpen{0.1mm}
\fmfset{curly_len}{0.4mm} \fmfset{arrow_len}{1mm}
\fmfset{dash_len}{1mm} \fmfset{dot_size}{1mm}
  \fmfsurroundn{v}{4}
  \fmf{phantom}{v1,w1}
  \fmf{phantom,tension=2}{v4,w2}
  \fmf{phantom}{v3,w3}
  \fmf{gluon,left}{w1,w2}
  \fmf{gluon,left,tension=0.5}{w2,w3}
  \fmf{gluon,left}{w3,w1}
  \fmf{gluon,left,tension=0.5}{w3,w2}
  \fmf{gluon}{w1,v1}
  \fmf{gluon}{v3,w3}
  \fmfv{decor.shape=circle,decor.filled=full,decor.size=0.6thick}{w1,w2,w3}
\end{fmfchar}}}

\newcommand{\qcdbggs}{\raisebox{-4mm}{\begin{fmfchar}(12,9) \fmfkeep{qcd2gg19} \fmfpen{0.1mm}
\fmfset{curly_len}{0.4mm} \fmfset{arrow_len}{1mm}
\fmfset{dash_len}{1mm} \fmfset{dot_size}{1mm}
  \fmfsurroundn{v}{4}
  \fmf{phantom}{v1,w1}
  \fmf{phantom,tension=2}{v2,w2}
  \fmf{phantom}{v3,w3}
  \fmf{gluon,left}{w3,w2}
  \fmf{gluon,left,tension=0.5}{w1,w2}
  \fmf{gluon,left}{w1,w3}
  \fmf{gluon,left,tension=0.5}{w2,w1}
  \fmf{gluon}{w1,v1}
  \fmf{gluon}{v3,w3}
  \fmfv{decor.shape=circle,decor.filled=full,decor.size=0.6thick}{w1,w2,w3}
\end{fmfchar}}}

\newcommand{\qcdbggt}{\raisebox{-2mm}{\begin{fmfchar}(12,9) \fmfkeep{qcd2gg20} \fmfpen{0.1mm}
\fmfset{curly_len}{0.4mm} \fmfset{arrow_len}{1mm}
\fmfset{dash_len}{1mm} \fmfset{dot_size}{1mm}
  \fmfsurroundn{v}{4}
  \fmf{phantom}{v1,w1}
  \fmf{phantom,tension=2}{v4,w2}
  \fmf{phantom}{v3,w3}
  \fmf{gluon,left}{w2,w3}
  \fmf{gluon,left,tension=0.5}{w1,w2}
  \fmf{gluon,left}{w3,w1}
  \fmf{gluon,left,tension=0.5}{w2,w1}
  \fmf{gluon}{w1,v1}
  \fmf{gluon}{v3,w3}
  \fmfv{decor.shape=circle,decor.filled=full,decor.size=0.6thick}{w1,w2,w3}
\end{fmfchar}}}

\newcommand{\qcdbggac}{\raisebox{-1mm}{\begin{fmfchar}(12,9) \fmfkeep{qcd2gg29} \fmfpen{0.1mm}
\fmfset{curly_len}{0.4mm} \fmfset{arrow_len}{1mm}
\fmfset{dash_len}{1mm} \fmfset{dot_size}{1mm}
   \fmfleft{bin,p1} 
   \fmfright{bout,p2} 
   \fmf{gluon}{bin,v,bout}
   \fmf{gluon,left,tension=0.3}{v,v1,v2,v}
   \fmf{gluon,left,tension=0.3}{v2,v1}
   \fmf{phantom,tension=0.5}{p1,v1}
   \fmf{phantom,tension=0.5}{v2,p2}
   \fmf{phantom,tension=0.5}{bin,v1}
   \fmf{phantom,tension=0.5}{v2,bout}
  \fmfv{decor.shape=circle,decor.filled=full,decor.size=0.6thick}{v,v1,v2}
\end{fmfchar}}}

\newcommand{\qcdbggad}{\raisebox{-1mm}{\begin{fmfchar}(12,9) \fmfkeep{qcd2gg30} \fmfpen{0.1mm}
\fmfset{curly_len}{0.4mm} \fmfset{arrow_len}{1mm}
\fmfset{dash_len}{1mm} \fmfset{dot_size}{1mm}
   \fmfleft{bin,p1} 
   \fmfright{bout,p2} 
   \fmf{gluon}{bin,v,bout}
   \fmf{gluon,left,tension=0.3}{v,v1}
   \fmf{gluon,left,tension=0.3}{v2,v}
   \fmf{fermion,left,tension=0.3}{v2,v1,v2}
   \fmf{phantom,tension=0.5}{p1,v1}
   \fmf{phantom,tension=0.5}{v2,p2}
   \fmf{phantom,tension=0.5}{bin,v1}
   \fmf{phantom,tension=0.5}{v2,bout}
  \fmfv{decor.shape=circle,decor.filled=full,decor.size=0.6thick}{v,v1,v2}
\end{fmfchar}}}

\newcommand{\qcdbggae}{\raisebox{-1mm}{\begin{fmfchar}(12,9) \fmfkeep{qcd2gg31} \fmfpen{0.1mm}
\fmfset{curly_len}{0.4mm} \fmfset{arrow_len}{1mm}
\fmfset{dash_len}{1mm} \fmfset{dot_size}{1mm}
   \fmfleft{bin,p1} 
   \fmfright{bout,p2} 
   \fmf{gluon}{bin,v,bout}
   \fmf{gluon,left,tension=0.3}{v,v1}
   \fmf{gluon,left,tension=0.3}{v2,v}
   \fmf{scalar,left,tension=0.3}{v2,v1,v2}
   \fmf{phantom,tension=0.5}{p1,v1}
   \fmf{phantom,tension=0.5}{v2,p2}
   \fmf{phantom,tension=0.5}{bin,v1}
   \fmf{phantom,tension=0.5}{v2,bout}
  \fmfv{decor.shape=circle,decor.filled=full,decor.size=0.6thick}{v,v1,v2}
\end{fmfchar}}}

\newcommand{\qcdbggab}{\raisebox{-1mm}{\begin{fmfchar}(12,9) \fmfkeep{qcd2gg28} \fmfpen{0.1mm}
\fmfset{curly_len}{0.4mm} \fmfset{arrow_len}{1mm}
\fmfset{dash_len}{1mm} \fmfset{dot_size}{1mm}
\fmfbottom{i,v1,v,v2,o}
\fmf{gluon}{i,v1,v,v,v2,o}\fmf{gluon,left,tension=0.3}{v1,v2}
\fmfv{decor.shape=circle,decor.filled=full,decor.size=0.6thick}{v1,v,v2}
\end{fmfchar}}}

\newcommand{\qcdbggaa}{\raisebox{-6mm}{\begin{fmfchar}(12,9) \fmfkeep{qcd2gg27} \fmfpen{0.1mm}
\fmfset{curly_len}{0.4mm} \fmfset{arrow_len}{1mm}
\fmfset{dash_len}{1mm} \fmfset{dot_size}{1mm} \fmftop{i,v1,v,v2,o}
 \fmf{gluon}{i,v1,v,v,v2,o}\fmf{gluon,left,tension=0.3}{v2,v1}
\fmfv{decor.shape=circle,decor.filled=full,decor.size=0.6thick}{v1,v,v2}
\end{fmfchar}}}

\newcommand{\qcdbggaf}{\raisebox{-5mm}{\begin{fmfchar}(12,9) \fmfkeep{qcd2gg32} \fmfpen{0.1mm}
\fmfset{curly_len}{0.4mm} \fmfset{arrow_len}{1mm}
\fmfset{dash_len}{1mm} \fmfset{dot_size}{1mm}
\fmfleft{bin}\fmfright{bout} \fmftop{tp} \fmf{gluon}{bin,v,bout}
\fmf{gluon,left,rubout,tension=0.1}{v,w,v}\fmf{phantom}{tp,w}\fmf{gluon,left,tension=1.3}{w,w}
  \fmfv{decor.shape=circle,decor.filled=full,decor.size=0.6thick}{v,w}
\end{fmfchar}}}

\newcommand{\qcdbggp}{\raisebox{-3mm}{\begin{fmfchar}(12,9) \fmfkeep{qcd2gg16} \fmfpen{0.1mm}
\fmfset{curly_len}{0.4mm} \fmfset{arrow_len}{1mm}
\fmfset{dash_len}{1mm} \fmfset{dot_size}{1mm}
\fmfleft{i}\fmfright{o} \fmf{gluon}{i,v1}\fmf{gluon}{v2,o}
\fmf{gluon,left,tension=0.3}{v1,v,v1}\fmf{gluon,left,tension=0.3}{v,v2,v}
  \fmfv{decor.shape=circle,decor.filled=full,decor.size=0.6thick}{v1,v,v2}
\end{fmfchar}}}

\newcommand{\qcdbggag}{\raisebox{-3mm}{\begin{fmfchar}(12,9) \fmfkeep{qcd2gg33} \fmfpen{0.1mm}
\fmfset{curly_len}{0.4mm} \fmfset{arrow_len}{1mm}
\fmfset{dash_len}{1mm} \fmfset{dot_size}{1mm}
\fmfleft{i}\fmfright{o} \fmf{gluon}{i,v1}\fmf{gluon}{v2,o}
\fmf{gluon,left,tension=0.2}{v1,v2,v1}\fmf{gluon,tension=0.2}{v1,v2}
  \fmfv{decor.shape=circle,decor.filled=full,decor.size=0.6thick}{v1,v2}
\end{fmfchar}}}


\newcommand{\qedzggb}{\raisebox{-0.5mm}{\begin{fmfchar}(4,4) \fmfkeep{qed0gg2} \fmfpen{0.1mm}
\fmfset{wiggly_len}{0.5mm}
  \fmfleft{bin} 
  \fmf{photon}{bin,bout}
  \fmfright{bout} 
\end{fmfchar}}}

\newcommand{\qedzff}{\raisebox{-0.5mm}{\begin{fmfchar}(4,4) \fmfkeep{qed0ff} \fmfpen{0.1mm}
\fmfset{wiggly_len}{0.5mm}\fmfset{arrow_len}{0.6mm}
\fmfset{dash_len}{0.7mm} \fmfset{dot_size}{1mm}
  \fmfleft{bin} 
  \fmf{fermion}{bin,bout}
  \fmfright{bout} 
\end{fmfchar}}}

\newcommand{\qedzgff}{\raisebox{-0.5mm}{\begin{fmfchar}(4,4) \fmfkeep{qed0gff} \fmfpen{0.1mm}
\fmfset{wiggly_len}{0.5mm} \fmfset{arrow_len}{0.6mm}
\fmfset{dash_len}{0.7mm} \fmfset{dot_size}{1mm}
\fmfleft{bin} 
  \fmf{photon}{bin,v}
  \fmf{fermion}{bout1,v,bout2}
  \fmfright{bout1,bout2} 
  \fmfv{decor.shape=circle,decor.filled=full,decor.size=0.6thick}{v}
\end{fmfchar}}}


\newcommand{\qedaffa}{\raisebox{-3mm}{\begin{fmfchar}(9,6) \fmfkeep{qed1ff1} \fmfpen{0.1mm}
\fmfset{wiggly_len}{0.6mm}\fmfset{arrow_len}{0.6mm}
\fmfset{dash_len}{0.4mm} \fmfset{dot_size}{1mm}
  \fmfleft{bin} 
  \fmf{fermion,tension=0.3}{v1,v2}
  \fmf{fermion}{bin,v1}
  \fmf{fermion}{v2,bout}
  \fmf{photon,left,tension=0.3}{v1,v2}
  \fmfright{bout} 
  \fmfv{decor.shape=circle,decor.filled=full,decor.size=0.6thick}{v1,v2}
\end{fmfchar}}}


\newcommand{\qedagga}{\raisebox{-2mm}{\begin{fmfchar}(10,6)
\fmfkeep{qed1gg1} \fmfpen{0.1mm} \fmfset{wiggly_len}{1mm}
\fmfset{arrow_len}{3mm}
  \fmfleft{bin} 
  \fmf{photon}{bin,v1}
  \fmf{photon}{v2,bout}
  \fmf{fermion,left,tension=.3}{v1,v2,v1}
  \fmfright{bout} 
  \fmfv{decor.shape=circle,decor.filled=full,decor.size=0.6thick}{v1,v2}
\end{fmfchar}}}

\newcommand{\qedagffa}{\raisebox{-2mm}{\begin{fmfchar}(9,6)
\fmfkeep{qed1gff1} \fmfpen{0.1mm} \fmfset{wiggly_len}{0.6mm}
\fmfset{arrow_len}{1mm} \fmfset{dash_len}{1mm}
  \fmfleft{bin} 
  \fmf{photon}{bin,v}
  \fmf{photon,tension=0.3}{v1,v2}
  \fmf{fermion,tension=0.3}{v2,v,v1}
  \fmf{fermion}{fout2,v2}
  \fmf{fermion}{v1,fout1}
  \fmfright{fout1,fout2} 
  \fmfv{decor.shape=circle,decor.filled=full,decor.size=0.6thick}{v,v1,v2}
\end{fmfchar}}}


\newcommand{\qedbggv}{\raisebox{-2mm}{\begin{fmfchar}(9,6) \fmfkeep{qed2gg22} \fmfpen{0.1mm}
\fmfset{wiggly_len}{0.6mm} \fmfset{arrow_len}{1mm}
\fmfset{dash_len}{1mm} \fmfset{dot_size}{1mm}
  \fmfsurroundn{v}{6}
  \fmf{phantom}{v1,w1}
  \fmf{phantom}{v2,w2}
  \fmf{phantom}{v3,w3}
  \fmf{phantom}{v4,w4}
  \fmf{phantom}{v5,w5}
  \fmf{phantom}{v6,w6}
  \fmf{fermion}{w3,w4,w5,w6,w1,w2}
  \fmf{fermion,left}{w2,w3}
  \fmf{photon,left}{w3,w2}
  \fmf{photon}{w1,v1}
  \fmf{photon}{v4,w4}
  \fmfv{decor.shape=circle,decor.filled=full,decor.size=0.6thick}{w1,w2,w3,w4}
\end{fmfchar}}}

\newcommand{\qedbggu}{\raisebox{-2mm}{\begin{fmfchar}(9,6) \fmfkeep{qed2gg21} \fmfpen{0.1mm}
\fmfset{wiggly_len}{0.6mm} \fmfset{arrow_len}{1mm}
\fmfset{dash_len}{1mm} \fmfset{dot_size}{1mm}
   \fmfleft{bin} 
   \fmfright{bout} 
  \fmfsurroundn{v}{6}
  \fmf{phantom}{v1,w1}
  \fmf{phantom}{v2,w2}
  \fmf{phantom}{v3,w3}
  \fmf{phantom}{v4,w4}
  \fmf{phantom}{v5,w5}
  \fmf{phantom}{v6,w6}
  \fmf{fermion}{w6,w1,w2,w3,w4,w5}
  \fmf{fermion,left}{w5,w6}
  \fmf{photon}{w6,w5}
  \fmf{photon}{w1,v1}
  \fmf{photon}{v4,w4}
  \fmfv{decor.shape=circle,decor.filled=full,decor.size=0.6thick}{w1,w5,w6,w4}
\end{fmfchar}}}

\newcommand{\qedbggw}{\raisebox{-2mm}{\begin{fmfchar}(9,6) \fmfkeep{qed2gg23} \fmfpen{0.1mm}
\fmfset{wiggly_len}{0.6mm} \fmfset{arrow_len}{1mm}
\fmfset{dash_len}{1mm} \fmfset{dot_size}{1mm}
  \fmfsurroundn{v}{4}
  \fmf{phantom}{v1,w1}
  \fmf{phantom,tension=2}{v2,w2}
  \fmf{phantom}{v3,w3}
  \fmf{phantom,tension=2}{v4,w4}
  \fmf{fermion}{w1,w2}
  \fmf{fermion}{w2,w3}
  \fmf{fermion}{w3,w4}
  \fmf{fermion}{w4,w1}
  \fmf{photon,tension=0.3}{w2,w4}
  \fmf{photon}{w1,v1}
  \fmf{photon}{v3,w3}
  \fmfv{decor.shape=circle,decor.filled=full,decor.size=0.6thick}{w1,w2,w3,w4}
\end{fmfchar}}}


\newcommand{\pppzggb}{\raisebox{-0.5mm}{\begin{fmfchar}(4,4) \fmfkeep{ppp0gg2} \fmfpen{0.1mm}
  \fmfleft{bin} 
  \fmf{plain}{bin,bout}
  \fmfright{bout} 
\end{fmfchar}}}

\newcommand{\pppzggg}{\raisebox{-0.5mm}{\begin{fmfchar}(4,4) \fmfkeep{ppp0ggg} \fmfpen{0.1mm}
 \fmfset{dot_size}{1mm}
\fmfleft{bin} 
  \fmf{plain}{bin,v}
  \fmf{plain,tension=0.2}{bout1,v,bout2}
  \fmfright{bout1,bout2} 
  \fmfv{decor.shape=circle,decor.filled=full,decor.size=0.6thick}{v}
\end{fmfchar}}}

\newcommand{\pppagga}{\raisebox{-0.5mm}{\begin{fmfchar}(4,4) \fmfkeep{ppp1gg1} \fmfpen{0.1mm}
  \fmfleft{bin} 
  \fmf{plain}{bin,v1}
  \fmf{plain}{v2,bout}
  \fmf{plain,tension=0.2,left}{v1,v2,v1}
 \fmfright{bout} 
    \fmfv{decor.shape=circle,decor.filled=full,decor.size=0.6thick}{v1,v2}
\end{fmfchar}}}

\newcommand{\pppaggga}{\raisebox{-0.5mm}{\begin{fmfchar}(3,3) \fmfkeep{ppp1gg1} \fmfpen{0.1mm}
  \fmfleft{bin} 
  \fmf{plain}{bin,v}
  \fmf{plain}{v1,bout1}
  \fmf{plain}{v2,bout2}
  \fmf{plain,tension=0.2}{v,v2,v1,v}
 \fmfright{bout1,bout2} 
    \fmfv{decor.shape=circle,decor.filled=full,decor.size=0.6thick}{v,v1,v2}
\end{fmfchar}}}

\newcommand{\pppbgga}{\raisebox{-0.5mm}{\begin{fmfchar}(3,3) \fmfkeep{ppp2gg1} \fmfpen{0.1mm}
  \fmfsurroundn{v}{4}
  \fmf{phantom}{v1,w1}
  \fmf{phantom,tension=2}{v2,w2}
  \fmf{phantom}{v3,w3}
  \fmf{phantom,tension=2}{v4,w4}
  \fmf{plain,tension=0.2}{w1,w2}
  \fmf{plain,tension=0.2}{w2,w3}
  \fmf{plain,tension=0.2}{w3,w4}
  \fmf{plain,tension=0.2}{w4,w1}
  \fmf{plain,tension=0.3}{w2,w4}
  \fmf{plain}{w1,v1}
  \fmf{plain}{v3,w3}
  \fmfv{decor.shape=circle,decor.filled=full,decor.size=0.6thick}{w1,w2,w3,w4}
\end{fmfchar}}}

\newcommand{\pppbggb}{\raisebox{-0.5mm}{\begin{fmfchar}(3,3) \fmfkeep{ppp2gg2} \fmfpen{0.1mm}
   \fmfleft{bin} 
   \fmfright{bout} 
   \fmf{plain}{bin,v1}
   \fmf{plain}{v2,bout}
   \fmf{plain,left,tension=0.1}{v1,v2}
   \fmf{plain,tension=0.1}{v2,v3}
   \fmf{plain,tension=0.1}{v4,v1}
   \fmf{plain,left,tension=0.05}{v4,v3,v4}
  \fmfv{decor.shape=circle,decor.filled=full,decor.size=0.6thick}{v1,v2,v3,v4}
\end{fmfchar}}}

\newcommand{\pppbggc}{\raisebox{-0.5mm}{\begin{fmfchar}(3,3) \fmfkeep{ppp2gg3} \fmfpen{0.1mm}
   \fmfleft{bin} 
   \fmfright{bout} 
   \fmf{plain}{bin,v1}
   \fmf{plain}{v2,bout}
   \fmf{plain,left,tension=0.1}{v2,v1}
   \fmf{plain,tension=0.1}{v2,v3}
   \fmf{plain,tension=0.1}{v4,v1}
   \fmf{plain,left,tension=0.05}{v4,v3,v4}
  \fmfv{decor.shape=circle,decor.filled=full,decor.size=0.6thick}{v1,v2,v3,v4}
\end{fmfchar}}}

\newcommand{\pppbggga}{\raisebox{-0.5mm}{\begin{fmfchar}(3,3) \fmfkeep{ppp2ggg1} \fmfpen{0.1mm}
  \fmfleft{bin} 
  \fmf{plain}{bin,v}
  \fmf{plain}{v1,bout1}
  \fmf{plain}{v2,bout2}
  \fmf{plain,tension=0.2}{v,w1,v1}
  \fmf{plain,tension=0.2}{v,w2,v2}
  \fmf{plain,tension=0.2}{w2,w1}
  \fmf{plain,tension=0.2}{v2,v1}
\fmfright{bout1,bout2} 
    \fmfv{decor.shape=circle,decor.filled=full,decor.size=0.6thick}{v,v1,v2,w1,w2}
\end{fmfchar}}}

\newcommand{\pppbgggb}{\raisebox{-0.5mm}{\begin{fmfchar}(3,3) \fmfkeep{ppp2ggg2} \fmfpen{0.1mm}
  \fmfleft{bin} 
  \fmf{plain}{bin,v}
  \fmf{plain}{v1,bout1}
  \fmf{plain}{v2,bout2}
  \fmf{plain,tension=0.2}{v,v1}
  \fmf{plain,tension=0.2}{v,w1,v2}
  \fmf{plain,tension=0.2}{w2,w1}
  \fmf{plain,tension=0.2}{v2,w2,v1}
\fmfright{bout1,bout2} 
    \fmfv{decor.shape=circle,decor.filled=full,decor.size=0.6thick}{v,v1,v2,w1,w2}
\end{fmfchar}}}

\newcommand{\pppbgggc}{\raisebox{-0.5mm}{\begin{fmfchar}(3,3) \fmfkeep{ppp2ggg3} \fmfpen{0.1mm}
  \fmfleft{bin} 
  \fmf{plain}{bin,v}
  \fmf{plain}{v1,bout1}
  \fmf{plain}{v2,bout2}
  \fmf{plain,tension=0.2}{v,w1,v1}
  \fmf{plain,tension=0.2}{v,v2}
  \fmf{plain,tension=0.2}{w2,w1}
  \fmf{plain,tension=0.2}{v2,w2,v1}
\fmfright{bout1,bout2} 
    \fmfv{decor.shape=circle,decor.filled=full,decor.size=0.6thick}{v,v1,v2,w1,w2}
\end{fmfchar}}}

\newcommand{\pppbgggd}{\raisebox{-0.5mm}{\begin{fmfchar}(3,3) \fmfkeep{ppp2ggg4} \fmfpen{0.1mm}
  \fmfleft{bin} 
  \fmf{plain}{bin,v}
  \fmf{plain}{v1,bout1}
  \fmf{plain}{v2,bout2}
  \fmf{plain,tension=0.2}{v,w1,v1}
  \fmf{plain,tension=0.2}{v,w2,v2}
  \fmf{plain,tension=0.2}{w1,v2}
  \fmf{plain,tension=0.2}{w2,v1}
\fmfright{bout1,bout2} 
    \fmfv{decor.shape=circle,decor.filled=full,decor.size=0.6thick}{v,v1,v2,w1,w2}
\end{fmfchar}}}

\newcommand{\pppbggge}{\raisebox{-0.5mm}{\begin{fmfchar}(3,3) \fmfkeep{ppp2ggg5} \fmfpen{0.1mm}
  \fmfleft{bin} 
  \fmf{plain}{bin,v}
  \fmf{plain}{v1,bout1}
  \fmf{plain}{v2,bout2}
  \fmf{plain,tension=0.2}{v,w1,w2,v1}
  \fmf{plain,tension=0.2}{v,v2}
  \fmf{plain,tension=0.2,left}{w2,w1}
  \fmf{plain,tension=0.2}{v2,v1}
\fmfright{bout1,bout2} 
    \fmfv{decor.shape=circle,decor.filled=full,decor.size=0.6thick}{v,v1,v2,w1,w2}
\end{fmfchar}}}

\newcommand{\pppbgggf}{\raisebox{-0.5mm}{\begin{fmfchar}(3,3) \fmfkeep{ppp2ggg6} \fmfpen{0.1mm}
  \fmfleft{bin} 
  \fmf{plain}{bin,v}
  \fmf{plain}{v1,bout1}
  \fmf{plain}{v2,bout2}
  \fmf{plain,tension=0.2}{v,v1}
  \fmf{plain,tension=0.2}{v,w1,w2,v2}
  \fmf{plain,tension=0.2,left}{w2,w1}
  \fmf{plain,tension=0.2}{v2,v1}
\fmfright{bout1,bout2} 
    \fmfv{decor.shape=circle,decor.filled=full,decor.size=0.6thick}{v,v1,v2,w1,w2}
\end{fmfchar}}}

\newcommand{\pppbgggg}{\raisebox{-0.5mm}{\begin{fmfchar}(3,3) \fmfkeep{ppp2ggg7} \fmfpen{0.1mm}
  \fmfleft{bin} 
  \fmf{plain}{bin,v}
  \fmf{plain}{v1,bout1}
  \fmf{plain}{v2,bout2}
  \fmf{plain,tension=0.2}{v,v1}
  \fmf{plain,tension=0.2}{v,v2}
  \fmf{plain,tension=0.2,left}{w2,w1}
  \fmf{plain,tension=0.2}{v2,w1,w2,v1}
\fmfright{bout1,bout2} 
    \fmfv{decor.shape=circle,decor.filled=full,decor.size=0.6thick}{v,v1,v2,w1,w2}
\end{fmfchar}}}